# Gate-Tunable and Thickness-dependent Electronic and Thermoelectric Transport in few-layer $MoS_2$

Morteza Kayyalha[1,2*], Jesse Maassen[2,3,4], Mark Lundstrom[2,4], Li Shi[5,6], Yong P. Chen[1,2,7,8*]

[1] Birck Nanotechnology Center, Purdue University, West Lafayette, IN, USA 47907

[2] School of Electrical and Computer Engineering, Purdue University, West Lafayette, IN, USA 47907

[3] Department of Physics and Atmospheric Science, Dalhousie University, Halifax, NS, Canada B3H 4R2

[4] Network for Computational Nanotechnology, Purdue University, West Lafayette, IN, USA 47907

[5] Department of Mechanical Engineering, University of Texas at Austin, Austin, TX, USA 78712

[6] Materials Science and Engineering Program, University of Texas at Austin, Austin, TX, USA 78712

[7] Department of Physics and Astronomy, Purdue University, West Lafayette, IN, USA 47907

[8] Purdue Quantum Center, Purdue University, West Lafayette, IN, USA 47907

[*] To whom correspondence should be addressed: mkayyalh@purdue.edu, yongchen@purdue.edu

**Abstract**

Over the past few years, there has been a growing interest in layered transition metal dichalcogenides (TMD) such as molybdenum disulfide ($MoS_2$). Most studies so far have focused on the electronic and optoelectronic properties of single-layer $MoS_2$, whose band structure features a direct bandgap, in sharp contrast to the indirect bandgap of thicker $MoS_2$. In this paper, we present a systematic study of the thickness-dependent electrical and *thermoelectric* properties of few-layer $MoS_2$. We observe that the electrical conductivity ($\sigma$) increases as we reduce the thickness of $MoS_2$ and peaks at about two layers, with six-time larger conductivity than our thickest sample (23-layer $MoS_2$). Using a back-gate voltage, we modulate the Fermi energy ($E_F$) of the sample where an increase in the Seebeck coefficient ($S$) is observed with decreasing gate voltage ($E_F$) towards the subthreshold (OFF state) of the device, reaching as large as $500\ \mu V/K$ in a four-layer $MoS_2$. While previous reports have focused on a single-layer $MoS_2$ and measured Seebeck coefficient in the OFF state, which has vanishing electrical conductivity and thermoelectric power factor ($PF = S^2\sigma$), we show that $MoS_2$-based devices in their ON state can have $PF$ as large as $> 50 \frac{\mu W}{cmK^2}$ in the two-layer sample. The $PF$ increases with decreasing thickness then drops abruptly from double-layer to single-layer $MoS_2$, a feature we suggest as due to a change in the energy dependence of the electron mean-free-path according to our theoretical calculation. Moreover, we show that care must be taken in thermoelectric measurements in the OFF state to avoid obtaining erroneously large Seebeck coefficients when the channel resistance is very high. Our study paves the way towards a more comprehensive examination of the thermoelectric performance of two-dimensional (2D) semiconductors.

## Introduction

Two-dimensional (2D) layered materials such as graphene and 2D semiconducting transition metal dichalcogenides (TMD) have recently gained a lot of attention due to their unique properties and potentials for applications in future electronics.[1–5] As a 2D TMD, molybdenum disulfide ($MoS_2$) is particularly promising because of its finite bandgap (1.8 eV in single layer compared to 1.2 eV in bulk), large $I_{on}/I_{off}$ ratio ($> 10^6$), good mobility and steep subthreshold slope ($\sim 75$ mV/dec) at room temperature.[6–9] Additionally, the band structure of $MoS_2$ shows a remarkable evolution with the thickness, transitioning from indirect to direct bandgap as the thickness decreases down to monolayer. This band structure change is expected to strongly affect the electrical and thermoelectric properties.[10,11]

While the main focus so far has been on the electrical and optoelectronic properties of TMDs[4,11–19], less attention has been paid to their thermoelectric properties. Seebeck coefficient ($S$) measurements are particularly sensitive to the particle/hole asymmetry and can provide unique insights into the electronic structure that may be more difficult to probe solely from standard electrical transport measurements.[20–23]

Previous studies have suggested that low dimensional systems can potentially achieve an improved thermoelectric $PF$ and figure of merit $ZT = \frac{PF}{\kappa}T = \frac{S^2\sigma}{\kappa}T$ , where $T$ is temperature and $\kappa$ is the thermal conductivity.[24–26] Researchers have also experimentally probed photo-thermoelectric and thermoelectric effects in single-layer $MoS_2$.[27,28] A few theories have predicted large $ZT$ values in $MoS_2$ and other TMD flakes of less than five layers in thickness at appropriate doping levels.[29,30] However, a careful experimental study of the thermoelectric transport in few-layer $MoS_2$ with tunable doping is needed to critically examine the thermoelectric performance in realistic $MoS_2$ materials.

In this letter, we present an experimental investigation of gate-tunable electrical and thermoelectric transport in single and few-layer $MoS_2$ field effect transistors (FET). Through application of the back-gate voltage ($V_G$), we are able to modulate the doping, electrical conductance, and Seebeck coefficient of $MoS_2$, where a notable enhancement of Seebeck coefficient is observed close to the threshold voltage ($V_{th}$). We also observe that the electrical conductivity increases as we reduce the thickness of few-layer $MoS_2$, reaching a peak at two layers. This enhancement in the electrical conductivity along with the fact that the Seebeck coefficient does not change significantly from 23 layers to two layers results in a six-time improvement in $PF$ of the two-layer sample compared to our thickest sample (23 layers). The gate voltage dependent electrical conductivity and Seebeck coefficient of single and double-layer $MoS_2$ are analyzed using a first principles-based approach, which indicates a stronger energy-dependent electron mean-free-path in the double layer resulting in higher Seebeck coefficient. Furthermore, we address a few issues in the Seebeck measurement of the back-gated semiconducting materials, especially in the subthreshold regime. These issues arise either from the large channel resistance of the device or the resistive coupling between the global back gate and the contact pads, and could result in unreliable Seebeck measurements. Our presented results, therefore, help to better understand the electrical and thermoelectric performance of $MoS_2$-based devices and also other TMDs and provide insight to their future applications as thermoelectric devices.

**Materials and Devices**

$MoS_2$ flakes with different thicknesses ($t$) ranging from single layer ($t \sim 0.65$ nm) to 23 layers ($t \sim 15$ nm) were exfoliated (from bulk $MoS_2$, obtained from 2dsemiconductors.com) using the standard scotch-tape technique and then transferred onto a degenerately doped silicon substrate with a 300 nm $SiO_2$ layer on top (see Figure 1a for a schematic). Electron-beam (e-beam) lithography followed by deposition of Al (70 nm) was utilized to make the contact probes, heater, and micro-thermometers. Previous studies report that low work function metals provide ohmic contacts to $MoS_2$.[31] We therefore choose Al because it has a low work function ($\sim 4.1 - 4.3$ eV), comparable with the electron affinity of $MoS_2$ (~ 4 eV)[14] and at the same time provides good adhesion for the following wire-bonding step. A combination of Atomic Force Microscopy (AFM) and Raman Spectroscopy was used to characterize the $MoS_2$ flakes. For thin (<~ 5 layers) $MoS_2$ flakes, the difference between the two dominant peaks ($E_{2g}^1$ and $A_{1g}$) in Raman spectra increases monotonically with the number of layers, and was used to determine the number of layers in this material.[32] For thicker flakes, AFM was employed to characterize them and measure their thickness (see

Appendix B). Our flakes are n-type with typical carrier mobility as measured by field effect around $20 - 60\ \frac{\text{cm}^2}{\text{Vs}}$ (see Appendix B).

Figure 1a shows a three dimensional schematic of a typical device used for Seebeck and electrical conductance measurements in our study (the optical image of a two-layer $MoS_2$ device is shown in the inset of Figure 1b). In this structure, two metal stripes ($R_1$ and $R_2$) simultaneously act as the source/drain contacts and micro-thermometers, while another metal line located adjacent to but not in direct contact with the flake acts as a micro-heater. In addition to the Seebeck measurement, this structure enables us to independently measure the two-probe and four-probe electrical conductance of the device. Moreover, the degenerately doped silicon substrate can be used as the back gate to tune $E_F$ or the carrier density in the $MoS_2$ channel.

**Measurement**

Electrical and thermoelectric transport measurements for our devices were performed in an evacuated cryostat, with pressure ~ $10^{-6}\ \text{Torr}$. Semiconducting field effect devices, especially in their subthreshold regime, have large channel resistances. This large resistance can become comparable with or larger than the input impedance of the measurement instruments. Therefore, careful consideration must be taken into account for the electrical and Seebeck measurements of these FETs. We use the voltage-biasing technique to measure the two-probe electrical conductance ($G_{2p}$) of our devices both in their ON and OFF regimes of operation. The four-probe electrical conductance ($G_{4p}$), however, was measured only in the ON state utilizing the standard current-biasing technique (for more details see Appendix B, Figure 10).

For a consistency check of our reported Seebeck coefficient, we used both DC and AC measurements and made sure that both techniques result in similar Seebeck coefficient values (for more details see Appendix C). In the DC configuration, a DC current is applied to the heater to create a temperature difference ($\Delta T$) across the channel, monitored by changes in the four-probe resistance of thermometers $R_1$ and $R_2$. A thermally induced DC voltage ($V_{thermal}$) between $R_1$ and $R_2$ is then measured using a Keithley 2182A nanovoltmeter and the resulting Seebeck coefficient is calculated from $S = -\frac{V_{thermal}}{\Delta T}$. When our devices enter their subthreshold regime, the channel resistance becomes very large. Therefore, a resistive coupling from the heater to the channel material (facilitated mostly through the contact pads) results in a deviation from the expected parabolic behavior in $V_{thermal}$ as a function of the heater current. This deviation becomes more pronounced as we go further into the subthreshold regime and can generate a spurious voltage signal that overwhelms the actual thermoelectric signal. Furthermore, the resistive coupling from the back gate to the channel and the small offset current from the nanovoltmeter will result in an offset voltage in $V_{thermal}$. This offset voltage, which is present even at zero heater current and is unrelated to the thermoelectric effect, as previously noted in other semiconducting channels such as Si MOSFETs,[33] could make Seebeck measurements further unreliable (for more details see Appendix C). When the device is in the ON state (the focus of this paper and where the Seebeck coefficient data presented below are measured), the channel resistance is small and as a result these spurious effects become rather small and insignificant.

In the AC configuration, a low frequency ($\omega$) AC heater current is applied to create a temperature difference ($\Delta T(2\omega)$, 90 deg phase shifted from the AC current) across the channel (between thermometers $R_1$ and $R_2$), monitored through changes in the four-probe resistance of $R_1$ and $R_2$(for more details see Appendix B). A thermally induced $2\omega$ voltage ($V_{thermal}(2\omega)$, 90 deg phase shifted from the AC current) is then monitored by a SRS 830 lock-in amplifier and the resulting Seebeck coefficient is calculated from $S = -\frac{V_{thermal}(2\omega)}{\Delta T(2\omega)}$. Our presented results in the main text are measured using the AC technique (with $\omega = 2\pi \times 5.117$ rad/sec) over a range of back-gate voltages where a reliable measurement could be performed (see Appendix C).

**Results and Discussion**

Figure 1b shows the output characteristic, the drain current ($I_D$) as a function of the drain voltage ($V_D$), for various back-gate voltages ($V_G$) of a representative field effect device (device #1) based on a two-layer $MoS_2$ ($t \sim 1.3$ nm) at room temperature. The linear behavior of $I_D$ versus $V_D$ is an indication of ohmic contacts. A room temperature transfer characteristic, $I_D$ versus $V_G$, of device #1 is also shown in Figure 1c. The increasing $I_D$ vs $V_G$ indicates n-type conduction. Two distinct regimes ("subthreshold/OFF" and "ON") of operation can be seen. In the subthreshold regime, the current increases exponentially as we increase $V_G$ (Figure 1c left axis) until $V_G$ moves above a threshold voltage (defined below) and the device turns ON. In the ON state and with the small drain voltage applied ($V_D = 400$ mV), the device is in its linear regime of operation and the current increases approximately linearly with increasing $V_G$ (Figure 1c right axis). The threshold voltage ($V_{th}$) is extracted by extrapolating the linear part of the transfer characteristic (for $V_G$ just above the subthreshold regime with exponential $I_D$-$V_G$ dependence) to zero $I_D$ as shown in Figure 1c.

Plotted in Figure 2a and b are the two-probe electrical conductance ($G_{2p}$) and Seebeck coefficient ($-S = |S|$) as functions of $V_G$ at six different temperatures ranging from 80 K to 300 K for device #1. The n-type behavior observed in the gate-dependent conductance (consistent with Figure 1c) of the device is in agreement with the negative sign of $S$ observed in the Seebeck measurement.

In Figure 2b, we plot $-S$ only in the ON state, where the $MoS_2$ channel is sufficiently conducting for $S$ to be measured reliably. For lower $V_G$, $E_F$ is lowered further into the band-gap and fewer charge carriers contribute to the transport. Even though we expect Seebeck coefficient to be significantly enhanced, the exceedingly large channel resistance can make Seebeck measurements unreliable (see Appendix C).

Figure 2c shows the four-probe electrical conductance ($G_{4p}$) of device #1 as a function of $V_G$ (above $-20$ V where the device is in the ON state) for six different temperatures. The n-type behavior, as seen by the increasing $G_{4p}$ with increasing $V_G$ (which raises $E_F$ further away from the mid-gap and toward/into the conduction band) is consistent with that seen in $G_{2p}$ and $-S$ (Figure 2a and b). On the other hand, $G_{4p}$, as a more intrinsic probe of the channel conduction compared to $G_{2p}$, reveals additional information in its temperature dependence. For $V_G > \sim 10$ V, $G_{4p}$ increases as we lower the temperature (metallic behavior), while for $V_G < \sim -10$ V, $G_{4p}$ decreases as we lower the temperature (insulating behavior). This transition from metallic to insulating behavior in $G_{4p}$ is further shown in Figure 2d where $G_{4p}$ is plotted as a function of $1/T$ for three different back-gate voltages. Such a transition, which has been previously observed in single and double layers of $MoS_2$,[15,34] is seen in most of our measured few-layer $MoS_2$ devices (see another example in 23-layer $MoS_2$ shown in Appendix B, Figure 12). We also observe that for the insulating regime (e.g. $V_G = -12$ and $-20$ V) the high-temperature part ($T \geq 120$ K) of $G_{4p}$ can be modeled by thermally activated transport (see Appendix B, Figure 11).[15] No metallic to insulating transition is seen in $G_{2p}$, which is likely dominated by the contact resistance due to Schottky barriers that become more significant as the temperature goes down. As a result, $G_{2p}$ decreases as we lower the temperature for all the back-gate voltages.

We now turn our attention to how the thickness (number of layers) of $MoS_2$ affects the electrical and thermoelectric transport properties in the ON state ($V_G > V_{th}$), where Seebeck measurements are reliable and appreciable thermoelectric $PF$ may be expected due to a larger electrical conductivity compared to the OFF state. For the ON state, the 2D charge carrier density inside the channel can be estimated using the parallel-plate capacitor model, $n_{2D} = \frac{C_G}{e}(V_G - V_{th})$, where $C_G$ is the $SiO_2$ capacitance per area and $e$ is the electron charge. As a result, Seebeck coefficient in various devices were compared at the same value of $V_G - V_{th}$, which corresponds to a certain 2D charge carrier density in the channel.

Figure 3a and b show the four-probe electrical conductivity ($\sigma_{4p} = G_{4p}\frac{L}{Wt}$, where $L$ and $W$ are the length and width of the $MoS_2$ channel, respectively) and Seebeck coefficient ($-S$) as functions of the back-gate voltage for devices with various channel thicknesses ($t$), respectively. The electrical conductivity shows an n-type behavior ($\sigma_{4p}$ increases with increasing $V_G$) for all our samples regardless of their thickness. Thickness dependent $\sigma_{4p}$ and $-S$ for different back-gate voltages are also presented in Figure 3c and d, respectively. As we change the channel thickness, the electrical conductivity ($\sigma_{4p}$) shows a maximum at two layers, while -$S$ has a peak at four layers. In particular, our results show as large as six-time improvement in $\sigma_{4p}$, as we reduce the channel thickness from 23 layers down to two layers. The dependence of $S$ on the channel thickness is much weaker compared to $\sigma_{4p}$ for devices thicker than single layer, whereas the single-layer flake gives notably smaller Seebeck coefficient ($\sim 200\ \mu V/K$) compared to thicker flakes ($\sim 400 - 500\ \mu V/K$).

Since using the back-gate voltage we can tune all our $MoS_2$ devices (even the 23-layer thick one) from ON state to OFF state, the back gate can modulate the carrier concentration of all the layers (the entire thickness) inside the $MoS_2$ channel (even though there could be some non-uniformity in the gating efficiency for different layers). We also know that the electrical conductivity $\sigma_{4p} = n_{3D}\mu e$ is proportional to the carrier mobility $\mu$ and 3D carrier concentration $n_{3D} = \frac{n_{2D}}{t}$, with the sheet carrier density $n_{2D} = \frac{C_G}{e}(V_G - V_{th})$, where $\mu$, $n_{3D}$ and $n_{2D}$ are understood as effective average values for the entire $MoS_2$ channel. Furthermore, in our comparison of the electrical conductivity (Figure 3a and b), $n_{2D}$ or $V_G - V_{th}$ is fixed. Therefore, we expect $n_{3D}$ ($= \frac{n_{2D}}{t}$) to be larger for the thinner flakes than bulk. Indeed, at a fixed $n_{2D}$, if we take into account the increase in $n_{3D}$(by 11.5 times going from 23 layers down to two layers) and variation in the carrier mobility (see Appendix B Figure 15), we estimate ~ 8 times improvement in $\sigma_{4p}$ of the double-layer $MoS_2$ compared to 23 layers, in reasonable agreement with the ~ 6 times improvement observed in our experiment. In other words, by reducing the layer thickness while keeping the electrostatic gating (thus $n_{2D}$) the same, we effectively increase the channel doping ($n_{3D}$), which in turn increases the electrical conductivity ($\sigma_{4p}$).

The thermoelectric $PF$ as functions of $V_G$ and number of layers are plotted in Figure 4a and 4b, respectively. The increasing $PF$ with increasing $V_G$ (doping) indicates our semiconducting channel is in the low-doping, non-degenerate regime, consistent with the finding that our field-effect Seebeck coefficient cannot be fitted from the conductivity by the Mott relation (see Figure 5a), which is derived for degenerate conductors ($E_F - E_C > k_B T$, where $k_B$ is the Boltzman constant). $PF$ vs. thickness also shows a peak for the two-layer device (device #1). As it can be seen from this figure and Figure 3, thickness-dependent $PF$ in $MoS_2$-based devices down to two layers is driven by the thickness dependence of the electrical conductivity of the device, since Seebeck coefficient does not vary strongly with the thickness. For the single-layer device, even though the electrical conductivity remains large and comparable to the two-layer device, $PF$ is much smaller mainly because of the smaller Seebeck coefficient. Our theoretical modeling, discussed later, provides insight into the nature of this measured difference in Seebeck coefficient between single and double-layer.

Large $PF$ (measured at the largest $V_G - V_{th}$ used in this experiment) of around $30 \frac{\mu\mathrm{W}}{\mathrm{cmK}^2}$, observed in the two-layer device here, is six times larger than $PF$ of around $5 \frac{\mu\mathrm{W}}{\mathrm{cmK}^2}$ observed in the 23-layer device. In another double-layer $MoS_2$, where we could apply larger $V_G$, a $PF \sim 53 \frac{\mu\mathrm{W}}{\mathrm{cmK}^2}$ was observed (see Appendix B Figure 16). Such a large $PF$ is notable given the highest $PF$ measured in the best bulk thermoelectric material $Bi_2Te_3$ is ~ $50 \frac{\mu\mathrm{W}}{\mathrm{cmK}^2}$.[35,36] Since the in-plane thermal conductivity ($\kappa$) of $MoS_2$ is relatively large, the $ZT$ would still be small ($\sim 0.05$, assuming reported values of $\sim 30 - 50 \frac{\mathrm{W}}{\mathrm{mK}}$ for $\kappa$[37–39]). On the other hand, we were unable to observe any peak in $PF$ within the range of the back-gate voltage used (< $70 - 100$ V, where the leakage current ($I_G$) starts to increase for higher $V_G$). Our results suggest that stronger gating towards more positive voltages (deeper in the ON state) may be needed to demonstrate the full thermoelectric potential of this material (see also Figure 5d and Appendix A for more details).

Utilizing first-principles density functional theory (DFT), we have calculated the band structure and the density of states, $DOS(E)$, for single- and double-layer $MoS_2$, see Appendix A, Figure 6.[26,40,41] Using the calculated electronic band dispersions combined with the Landauer approach [26,40–43], we compute the Seebeck coefficient ($S$) and electrical conductivity ($\sigma_{4p}$) at $T = 300$ K as functions of $E_F - E_C$, the Fermi energy ($E_F$) relative to the bottom of the conduction band ($E_C$). We assume a power-law energy dependent electron mean-free-path for backscattering $\lambda = \lambda_0 \left(\frac{E-E_C}{k_B T}\right)^r$, where $\lambda_0$ and $r$ are two fitting parameters independent of energy (here the mean-free-path for backscattering is the distance travelled along the transport direction ($x$) before scattering changes the sign of the momentum along that direction, $k_x$). We also calculate the relationship between $E_F - E_C$ and the carrier density. We can relate the back-gate voltage to the position of Fermi energy ($E_F - E_C$) and the corresponding 2D carrier concentration by fitting the experimental gate-dependent Seebeck coefficient (see Appendix A Figure 8) and the $V_{th}$ obtained from such fits are close to those extracted from experimental $I_D - V_G$ curves (e.g. Figure 1c). We have compared the calculated $-S$ and $\sigma_{4p}$ vs. $V_G - V_{th}$ for single and double-layer $MoS_2$ using various different trial r values to the experimental results, and the best fits (plotted in Figure 5a,b) are achieved using a constant electron backscattering mean-free-path ($\lambda_0^{1L} = 0.7$ nm, $r^{1L} = 0$) for single-layer $MoS_2$, and $\lambda^{2L} = \lambda_0^{2L} \frac{E-E_C}{k_B T}$ ($\lambda_0^{2L} = 0.8$ nm, $r^{2L} = 1$) for double-layer $MoS_2$. The $r^{2L} = 1$ value for double-layer $MoS_2$ implies that the average mean-free-path increases with increasing $V_G$ (see Figure 8), and suggests that ionized impurity scatterings may be playing a role.[44] We note that this difference in the energy-dependence of $\lambda$ is important in explaining the smaller Seebeck coefficient observed in our single layer $MoS_2$ compared to double layer.

From our theoretical analysis, we can also relate the back-gate voltage to the position of Fermi energy with respect to the bottom of the conduction band ($E_F - E_C$) and the corresponding 2D carrier concentration ($n_{2D}$ see Figure 8). For the range of applied back-gate voltages in Figure 4, we find that the $n_{2D}$ varies between $1.2\times10^{12}$ cm$^{-2}$ (corresponding to $E_F - E_C \sim -46$ meV) and $5.5\times10^{12}$ cm$^{-2}$ (corresponding to $E_F - E_C \sim -2$ meV) for the single layer, and between $8.9\times10^{11}$ cm$^{-2}$ (corresponding to $E_F - E_C \sim -78$ meV) and $5.2\times10^{12}$ cm$^{-2}$ (corresponding to $E_F - E_C \sim -30$ meV) for the double-layer $MoS_2$. We note that $E_F - E_C$ for both devices is always negative, thus the Mott formula (see equation A1) cannot be used here. To demonstrate this fact, we have plotted $-S$ vs. $V_G - V_{th}$ (black dashed line) calculated from the Mott relation using the measured electrical conductivity for the single-layer device in Figure 5a.

Figure 5c shows the dependence of $-S$ on the four-probe conductivity. As expected, Seebeck coefficient increases with a decreasing conductivity. Using the extracted relationship between the back-gate voltage and Fermi level position, we plot the measured and calculated thermoelectric $PF$ vs. $E_F - E_C$ in Figure 5d. We predict a peak of $\sim 95 \frac{\mu W}{cmK^2}$ around $E_F - E_C \sim 82$ meV in the calculated thermoelectric $PF$ of the double-layer $MoS_2$. Our theoretical model can also provide the electronic component of the thermal conductivity using the experimentally-calibrated energy-dependent mean-free-paths for single and double-layer $MoS_2$. Our results show that the maximum values (for the range of applied back-gate voltages) of electronic thermal conductivity are $\sim 0.1\,\frac{\mathrm{W}}{\mathrm{mK}}$ and $\sim 0.14\,\frac{\mathrm{W}}{\mathrm{mK}}$ for single and double layer, respectively. Thus, the electronic thermal conductivity is expected to be much smaller than the lattice thermal conductivity ($\sim 30-50\,\frac{\mathrm{W}}{\mathrm{mK}}$), which indicates that reducing the lattice thermal conductivity would help increase $ZT$ . Considering the theoretical maximum power factor and assuming a thermal conductivity of $\sim 30-50\,\frac{\mathrm{W}}{\mathrm{mK}}$, we obtain a maximum $ZT \sim 0.1$ for the double-layer device. Lastly, our analysis suggests that the larger thermoelectric $PF$ of double-layer $MoS_2$ compared to single-layer $MoS_2$ is a result of the stronger energy dependence of the electron mean-free-path which significantly increases the Seebeck coefficient.

**Conclusion**

In conclusion, the gate modulated electrical conductivity and Seebeck coefficient were measured in $MoS_2$ flakes with different thicknesses. We have observed the largest thermoelectric power factor ($PF$) in two-layer $MoS_2$, about six times improved compared to the thickest (23-layer) $MoS_2$ film. This increase in $PF$ stems from a larger $\sigma_{4p}$ with comparable $S$ to that of thicker flakes. We also explained the significant drop in Seebeck coefficient with the single-layer $MoS_2$ compared to double-layer $MoS_2$, which in turn results in smaller thermoelectric $PF$, as arising from different energy dependences of $\lambda$ (constant for the single layer and linear for the double layer). Furthermore, from our fit of $S$ and $\sigma_{4p}$ we predict the maximum power factor of $\sim 95\,\frac{\mu\mathrm{W}}{\mathrm{cmK}^2}$, corresponding to the maximum $ZT \sim 0.1$, in our bi-layer $MoS_2$. We have also pointed out that the large channel resistance of the back-gated FETs in the subthreshold regime could make Seebeck measurements unreliable and result in erroneously large Seebeck coefficient values. Our observations bring new insights to understanding of the electronic and thermoelectric properties of $MoS_2$ and will help to explore the possibility of using $MoS_2$ and other TMDs in the future thermoelectric applications.

**Acknowledgement**

We acknowledge partial support from DARPA (Grant N66001-11-1-4107). J. M. acknowledges support from NSERC. We also acknowledge helpful discussions with Helin Cao and Jiuning Hu. We also thank Aida Ebrahimi for her help with COMSOL simulation.

## Appendix A: Theoretical modeling of thermoelectric properties

### A.1 Mott Relation

The Mott formula, which is derived for degenerately doped materials, can be described as[20,23]:

$$S = -\frac{\pi^2 k_B^2 T}{3e} \frac{1}{\sigma_{4p}} \frac{d\sigma_{4p}}{dV_G} \frac{dV_G}{dE} |_{E=E_F} \tag{A1}$$

For n-type single-layer $MoS_2$ with a parabolic band dispersion ($E = \frac{\hbar^2 k^2}{2m^*}$) and approximate spin degeneracy of $g = 2$ and valley degeneracy of $g_v = 2$, we obtain $\frac{dV_G}{dE} = \frac{2em^*}{C_G \pi \hbar^2}$. Therefore, we have:

$$S = -\frac{2\pi m^* k_B^2 T}{3C_G \hbar^2} \frac{1}{\sigma_{4p}} \frac{d\sigma_{4p}}{dV_G} \tag{A2}$$

The Seebeck coefficient of single-layer $MoS_2$ calculated from the Mott formula (equation A2) is plotted together with the experimental $S$ in Figure 5a. Since our single-layer $MoS_2$ is not degenerately doped, the Mott relation does *not* give a good estimate for $S$ vs. $V_G - V_{th}$.

### A.2. Landauer Formalism

The in-plane thermoelectric properties of single- and double-layer $MoS_2$ are calculated using the Landauer transport formalism, which is equivalent to solving the Boltzmann equation in the case of diffusive transport [26,40–43]. Here, we will briefly describe our approach to calculate the Seebeck coefficient and electrical conductivity using the full band dispersions obtained from the first-principles density functional theory (DFT). More elaborate discussion of our method can be found elsewhere [40,41].

Figure 6 shows the electronic dispersion for single-layer and double-layer $MoS_2$ calculated by density functional theory (DFT). Our calculated band structure shows that single-layer $MoS_2$ is a direct gap semiconductor with a 1.68 eV band gap and double-layer $MoS_2$ is an indirect gap semiconductor with a 1.34 eV band gap. The electronic states were calculated using the DFT-based VASP simulation software [45,46], which uses a plane-wave basis to expand the eigenfunctions (energy cutoff of 400 eV for single-layer and double-layer $MoS_2$) and the projector augmented-wave method to treat the atomic cores. The PBE flavor of the generalized gradient approximation (GGA) was employed, along with the optimized lattice constants taken from Ref. [47]. A 7×7×1 Monkhorst-Pack-generated k-grid was utilized for the self-consistent charge density calculations.

We model the thermoelectric properties of single and double-layer $MoS_2$ using Landauer formalism. In this approach, the four-probe sheet conductance ($G_{sheet}$) and Seebeck coefficient ($S$) can be expressed as:

$$G_{sheet} = \frac{2e^2}{h}\int_{-\infty}^{\infty} M_e(E)\lambda(E)\left[-\frac{\partial f(E)}{\partial E}\right]dE \qquad \text{(A3)}$$

$$S = -\frac{1}{eT}\frac{\int_{-\infty}^{\infty}(E-E_F)M_e(E)\lambda(E)\left[-\frac{\partial f(E)}{\partial E}\right]dE}{\int_{-\infty}^{\infty} M_e(E)\lambda(E)\left[-\frac{\partial f(E)}{\partial E}\right]dE} \qquad \text{(A4)}$$

where $e = 1.6\times10^{-19}$ C is the magnitude of the electron charge, $h$ is the Planck's constant, $T$ is the temperature, $M_e(E)$ is the number of modes, $\lambda(E)$ is the electronic mean-free-path for backscattering, $E_F$ is the Fermi energy, and $f(E) = \frac{1}{exp(\frac{E-E_F}{k_B})+1}$ is the Fermi distribution function.

$M_e(E)$ depends only on the calculated electronic dispersion of single-layer and double-layer $MoS_2$[26], which we extract using the LanTraP tool[47]. For $\lambda(E)$, we assume an expression of the form $\lambda = \lambda_0\left(\frac{E-E_C}{k_BT}\right)^r$, where $\lambda_0$ and $r$ are two fitting parameters independent of energy. With $\lambda$, we can compute the sheet conductance ($G_{sheet}$) and Seebeck coefficient ($S$) as functions of 2D carrier concentration ($n_{2D} = \int_{E_C}^{\infty} DOS(E)f(E)dE$) from equations A3 and A4, respectively. Furthermore, using the parallel-plate capacitor model ($n_{2D} = \frac{C_G}{e}(V_G - V_{th}^0)$) we can relate $n_{2D}$ and $V_G$. Here $V_{th}^0$ is an intrinsic threshold voltage that we obtain from our fitting (determined from the procedure described below) and can be slightly different from the value ($V_{th}$) extracted from the $I_D - V_G$ curve (e.g., Figure 1c). The procedure to calculate the fitting parameters is as follows. First we fit $S$ vs. $V_G$ to extract $r$ and $V_{th}^0$, note that $\lambda_0$ cancels out in the expression (equation A4) for $S$. Second, we fit $G_{sheet}$ vs. $V_G$ to determine $\lambda_0$. We only vary $r$ in increments 0.5 instead of continuously to obtain the best possible fit. Through our fitting procedure, we find that for single-layer $MoS_2$ the optimal parameters are $\lambda_0^{1L} = 0.7$ nm, $r^{1L} = 0$ and for double-layer $MoS_2$, the fitted parameters are $\lambda_0^{2L} = 0.8$ nm, $r^{2L} = 1$. Additionally, from fitted $V_{th}^0$ we obtain $E_F - E_C = -70$ meV with $n_{2D}$= 4.8×$10^{11}$cm$^{-2}$ for single-layer, and $E_F - E_C = -120$ meV with $n_{2D}$= 1.7×$10^{11}$cm$^{-2}$ for double-layer $MoS_2$, all evaluated at $V_G = V_{th}$. Figure 7a and b plot $-S$ and $G_{sheet}$ as functions of $E_F - E_C$. From our calculations, thermoelectric $PF$ of the double-layer $MoS_2$ shows a predicted peak at $E_F - E_C \sim 82$ meV, corresponding to $V_G - V_{th} \sim 1921$ V.

Figure 8a and b plot the calculated 2D carrier concentration ($n_{2D}$) vs. $E_F - E_C$ and $E_F - E_C$ vs. $V_G - V_{th}$, respectively. Figure A8c and d plot calculated mobility (defined as $\mu = \frac{G_{sheet}}{ne}$) and average mean-free-path ($\lambda_{average} = \frac{\int_{-\infty}^{\infty} \lambda(E) M_e(E) \left[-\frac{\partial f(E)}{\partial E}\right] dE}{\int_{-\infty}^{\infty} M_e(E) \left[-\frac{\partial f(E)}{\partial E}\right] dE}$)[48] vs. $V_G - V_{th}$ for single and double-layer $MoS_2$. As it can be seen, for the double layer sample, the average mean-free-path increases with increasing $V_G$, while for the single layer sample, the average mean free path is constant ($\lambda_0^{1L} = 0.7$ nm, since $r^{1L} = 0$).

## Appendix B: Additional experimental results

### B1. AFM and Raman Characterization

In order to confirm the thickness of $MoS_2$ flakes up to four layers, Raman Spectroscopy (Horiba XploRA Raman spectrometer with 532 nm laser light) was used. Figure 9a shows the results for single to four layers of $MoS_2$. The differences between the major two peaks ($E_{2g}^1$ and $A_{1g}$) are 18.5, 21.8, and 24 cm$^{-1}$ corresponding to 1, 2, and 4 layers, respectively.[32] AFM was also performed to characterize thicker flakes. The AFM result for a six-layer device is presented in Figure 9b.

### B2. Electrical and thermoelectric transport measurements

Figure B10a and b show our measurement set-up for four-probe and two-probe electrical measurements, respectively. In semiconducting FETs, especially in their OFF state, the channel resistance can become comparable to or larger than the instrument impedance. Therefore, a normal two-probe or four-probe current-biasing technique could result in unreliable measurements of the electrical conductance in the OFF state (where the voltmeter could shunt away a notable part of the current). Additionally, applying a small current (as small as 100 nA) in the OFF state where the channel resistance is large will result in a significant voltage drop ($V_D$) across the channel. This large $V_D$ will put the device in its high-field region and results in inaccurate conductance measurement. Therefore, we use the voltage-biasing technique to measure the two-probe conductance of the device in both the ON and OFF states. In this way, we make sure that having a large channel resistance will not affect our measurement. At the same time, $V_D$ is always small ($V_D \sim 100 - 400$ mV) to ensure that we are in the low-field region.

We used the AC current-biasing technique for the four-probe electrical measurement. This measurement was performed in the ON state where the channel is sufficiently conductive for $G_{4p}$ to be measured reliably.

The temperature dependence of $G_{4p}$ for device #1 (a 2-layer $MoS_2$, studied in Figures 1 and 2) is presented in Figure 11a for the insulating part of Figure 2c. We observe that for the high-temperature part ($T \geq 120$ K), $G_{4p}$ can be modeled by thermally activated transport[15]:

$$G = G_0 e^{-\frac{E_a}{k_B T}}$$

where $E_a$ is the activation energy, $k_B$ is the Boltzmann constant, and $G_0$ is a parameter that can be extracted from the fitting. The thermal activation model, however, cannot be used at temperatures below $T < 120$ K. At these low temperatures, transport might be dominated by a variable range hopping through localized states.[15,49,50] Figure 11b shows $E_a$ as a function of $V_G$.

Figure 12 shows $G_{4p}$ as a function of $V_G$ in 23-layer ($t \sim 15$ nm) $MoS_2$ with Al contacts (70 nm thick). We observe a metal to insulating transition tuned by $V_G$ in this sample even with its relatively thick channel.

Figure 13 depicts the details of our Seebeck measurement set-up.

In the AC configuration, as we apply a low frequency AC current to the heater ($I_{Heater}$, with a SR830 lock-in amplifier), a temperature difference is built up across the device, causing a thermally-induced voltage ($V_{thermal}$, at frequency of $2\omega$ and 90 deg phase shifted from the AC current) between the two voltage probes ($R_1$ and $R_2$, which are also used as thermometers). The temperature rises ($\delta T_1(2\omega)$ and $\delta T_2(2\omega)$) at $R_1$ and $R_2$ are measured through changes in the four-probe resistance of each thermometer ($\Delta R$). These temperature rises as well as the thermoelectric voltage $V_{thermal}$ are all found to be proportional to $I_{Heater}^2$ (Figure 14a and b, note all quantities are lock-in detected RMS values). The resulting Seebeck coefficient is then calculated through $S = -\frac{V_{thermal}}{\Delta T}$, where $\Delta T(2\omega) = \delta T_1(2\omega) - \delta T_2(2\omega)$ is the temperature difference across the channel.

$\Delta R$ in our thermometers ($R_1$ and $R_2$) is measured by applying a DC current ($I_{DC} \sim 100 - 200$ μA) to each thermometer and monitoring the voltage drop across the thermometer at $2\omega$ frequency ($\Delta V\ (2\omega)$) while the AC heater current ($I(\omega)$) is gradually raised. The change in the resistance of each thermometer can then be calculated by $\Delta R(2\omega) = \frac{\Delta V(2\omega)}{I_{DC}}$. We have also calibrated the temperature coefficient ($\alpha = \frac{\Delta R}{R\delta T}$) of each thermometer separately by monitoring $R_1$ and $R_2$ (measured by standard 4-probe method using the lock-in amplifier) as we varied the temperature ($T$) of our samples using a heater. We can then extract $\delta T$ for each thermometer as $\delta T = \frac{\Delta R}{R\alpha}$. Considering the geometry of the heater and thermometers (heater length is larger than both thermometers) in our devices, we believe that the temperature along each thermometer is nearly uniform.[51]

The temperature rise ($\delta T$, measured in the AC mode) for device #1 is plotted for both thermometers as a function of $I_{Heater}$ in Figure 14a. Figure 14b also illustrates $V_{thermal}$ (measured in the AC mode) across the channel in the ON state for two different back-gate voltages, as a function of $I_{Heater}$. As expected, both $\delta T$ and $V_{thermal}$ increase in a parabolic manner as we increase $I_{Heater}$. Figure 14c illustrates the temperature profile (caused by the Joule heating) in our double-layer $MoS_2$ structure (Figure 1b) calculated from finite-element simulation (using the software package COMSOL). The simulated temperature difference is 12% lower than that we measure experimentally.

In the DC configuration, the thermally induced voltage was measured with a Keithley 2182A nanovoltmeter, which has an input impedance > 10 GΩ. This will help reduce the uncertainty of the measured voltage, especially in the subthreshold regime of operation. However, in this DC approach due to additional problems that will be discussed in Appendix C, the heater current must be swept at each $V_G$ in order to make sure that the open-circuit voltage ($V_{open-circuit}$) is indeed caused by the thermoelectric effects (which should show parabolic behavior for $V_{open-circuit}$ as a function of $I_{Heater}$).

Figure 15a shows the four-probe sheet conductance or 2D conductivity ($G_{sheet} = G_{4p}\frac{L}{W}$) as a function of $V_G - V_{th}$ for devices with various channel thicknesses. Thickness-dependent $G_{sheet}$ at different back-gate voltages are also presented in Figure 15b. Figure 15c plots $\sigma_{4p}$ as a function of the layer thickness when the back gate is grounded ($V_G = 0$ V). Figure 15d (right axis) shows the highest mobility ($\mu = \frac{1}{C_G}\frac{dG_{sheet}}{dV_G}$) as a function of thickness. As we expect from $G_{sheet} = n_{2D}\mu e$ (where $n_{2D}$ is the 2D carrier concentration and $e = 1.6\times10^{-19}$ C is the magnitude of the electron charge) at a fixed $n_{2D}$ (fixed $V_G - V_{th}$), $G_{sheet}$ vs. thickness ($t$) shows a similar trend as that of mobility ($\mu$) vs. $t$. $G_{sheet}$vs. thickness ($t$) at $V_G = 0$ V is also plotted in Figure 15d (left axis). Figure 15e and f depict $-S$ and $PF$ vs. thickness at $V_G = 0$ V, respectively.

Figure 16a represents the four-probe electrical conductivity (left) and Seebeck coefficient (right) of another double-layer $MoS_2$ sample fabricated using Al contacts. Since we could apply larger positive $V_G$ in this device, we were able to observe larger $PF$ ($\sim 53\ \frac{\mu W}{cmK^2}$).

## Appendix C: Notes on the electrical and thermoelectric measurements of a semiconducting channel

In the OFF state, the channel resistance becomes comparable with or larger than the input resistance of the voltmeter. In this case, having a small gate leakage current, which is normal for these back-gated devices, or a small leakage current of the voltmeter itself might become problematic.

Here, we perform extensive AC and DC measurements in order to identify whether the measured $V_{open-circuit}$ is in fact due to the thermoelectric effects from the channel material (e.g. $V_{open-circuit} = V_{thermal}$) or it is simply a result of instrumental or experimental artifacts.

For this investigation, another $MoS_2$ device with only one heater and two microthermometers was fabricated (inset of Figure 17a). Figure 17a plots the two-probe resistance (left) and conductance (right) of the channel, measured by applying a constant $V_D$ of 100 mV. The threshold voltage ($V_{th}$) of the device is around $-1.5$ V.

### C.1. DC measurement

In the DC mode and in order to measure $V_{open-circuit}$, we use a Keithley 2182A nanovoltmeter, with more than 10 GΩ input resistance. The heater current and the back-gate voltage are supplied by a Keithley 2162A source meter.

Figure 17b shows the resulting $V_{open-circuit}$ as a function of $I_{Heater}$ for two different back-gate voltages of $-40$ and $-30$ V. As it can be seen, $V_{open-circuit}$ vs. $I_{Heater}$ does not behave in a parabolic fashion that should be expected for a thermally-induced voltage, and there is also an offset voltage at zero heater current. Both these phenomena can be explained considering that the device resistance is very large in the OFF region.

In the OFF state, a resistive coupling (through the 300nm-thick $SiO_2$) from the heater to the channel material (facilitated mostly through the contact pads as shown in Figure 18) results in a deviation from the parabolic behavior in the open-circuit voltage as the heater current changes from $-4$ to 4 mA (Figure 17b), while the resistive coupling from the back gate to the channel results in a constant offset voltage (even at zero heater current) in $V_{open-circuit}$ (e.g., notable in the blue curve in Figure 17c). For example, a gate-oxide resistance of around 0.3 TΩ and a channel resistance of around 10 MΩ can be calculated from our data for the back-gate voltage of $-10$ V. This will result in an offset voltage (through the resistive coupling) of around $300\ \mu V$ ($\sim 10\ V \times \frac{10\ M\Omega}{0.3\ T\Omega}$), which is on the same order as that observed in Figure 17c (blue curve).

It should be mentioned that nanovoltmeter offset current ($60 - 100$ pA) is also partially responsible for this constant offset voltage.[33] As we get closer to the onset of the ON state, more parabolic behavior is observed in the open-circuit voltage (Figure 17c, blue curve) and finally it becomes completely parabolic once we are inside the ON regime of the FET (Figure 17c, red curve). Gate-dependent open-circuit voltage in the OFF state of the device for three different heater currents is shown in Figure 17d. As it can be seen from this figure, just by looking at the open-circuit voltage, when the heater is ON, one can report Seebeck coefficient values as large as $10\ V/K$ or more. However, thermoelectric effects are not responsible for this open-circuit voltage. In order to give an estimate of how much these spurious effects contribute to the measured signal in one special case, we have fitted the blue line ($V_G = -10$ V) in Figure 17(c) to a second-degree polynomial ($a_0 + a_1 I_{Heater} + a_2 I_{Heater}^2$). We note that the constant term ($a_0$) corresponds to the contribution of the resistive coupling from the back-gate voltage and also nano-voltmeter offset current. The linear term, $a_1 I_{Heater}$, indicates the contribution of coupling from the heater pads to the channel (through the back gate). And finally the second-order term ($a_2 I_{Heater}^2$) is the contribution of the thermoelectric effects in the measured signal. Using this fitting, we obtain $a_0 = -312\ \mu V$, $a_1 = 7.77\ \mu V/mA$, and $a_2 = -14.35\ \mu V/mA^2$. We, therefore, find that the constant term ($a_0$) is 136 % and the linear term ($a_1 I_{Heater}$) is 13.5% of the actual thermoelectric signal ($a_2 I_{Heater}^2 = -229.6 \frac{\mu V}{K}$) when $I_{Heater} = 4$ mA.

In order to further investigate this issue in the OFF state, we used an Agilent 4145A Semiconductor Parameter Analyzer (SPA) with more than $10^{13}\Omega$ input resistance. In this case, all other measurement units were disconnected from the device and the device was only connected to the Source Measurement Units (SMU) of our SPA. Figure 19a and b illustrate the results from this experiment. As it can be seen, behaviors observed for $V_{open-circuit}$ as a function of $I_{Heater}$ (Figure 19a) and $V_G$ (Figure 19b) are similar to our previous measurement with the Keithley 2182A.

### C.2. AC measurement

We designed two different experiments in the AC mode. In the first experiment, a SR830 lock-in amplifier (input impedance $\sim 10\ M\Omega$) is directly used to measure $V_{open-circuit}$ across the channel. In this case, we observe a strong frequency-dependence for Seebeck coefficient, especially when the device enters its OFF state (Figure 20a red and green dashed lines).

In the second experiment, a SR560 Pre-Amplifier (input resistance $> 100\ M\Omega$) is used to reduce the loading effects on the lock-in amplifier. As it can be observed from Figure 20a, measured Seebeck coefficient values are independent of the measurement frequency for $f < 20$ Hz.

In both experiments, when the device enters the OFF state, the phase shift ($\theta$, between $I_{Heater}$ and $V_{open-circuit}$) deviates from 90 deg. This can be seen in Figure 20b, where $\theta$ is plotted as a function of $V_G$ when the lock-in amplifier is directly used. As a result, the in-phase component (meaning 90 deg phase shifted from $I_{Heater}$) of the lock-in amplifier is not a good measurement of $V_{open-circuit}$ and the AC measurement becomes unreliable. This behavior is observed in more than 20 samples, regardless of their channel thickness. The frequency dependence and substantial out-of-phase component seen in the measured signal likely occurs when the input impedance of the lock-in amplifier becomes comparable or even smaller than the device in the OFF state. Using a preamplifier helps to reduce this loading effect and thus the phase shift observed is similar to the case where we use directly a lock-in amplifier with $f = 5.117$ Hz.

As it is shown in Figure 20a, both the low frequency AC and DC measurements result in similar values of Seebeck coefficient. Therefore, we picked the low frequency of 5.117 Hz for the thickness-dependent thermoelectric measurement in $MoS_2$-based FETs.

In conclusion, the AC technique is limited by the input impedance of the lock-in amplifier. Using a preamplifier would help to reduce the loading effects, however in our case the input impedance of our preamplifier was limited to 100 MΩ which was lower than the channel resistance in the OFF state ($V_G < V_{th}$). As a result, we could perform reliable Seebeck measurements only in the ON state ($V_G > V_{th}$). Our presented results in the main text are measured using the AC technique (with $f = 5.117$ Hz) over a range of back-gate voltages where a reliable measurement could be performed.

In the DC mode, in addition to the input impedance of the nanovoltmeter, the leakage currents (either from the nanovoltmeter or the back-gate voltage) and resistive coupling from the pads to the highly doped silicon back gate impose difficulties in Seebeck measurements in the OFF state of the device. Therefore, at each gate voltage the heater current must be swept to make sure the measured voltage is indeed caused by theremoelectric effects.

## Figures

Figure 1. (a) Schematic (not to scale) of a typical device used for thermoelectric and electrical measurements. (b) Room temperature output characteristic ($I_D - V_D$) of device #1 (two-layer $MoS_2$) for various back-gate voltages ($V_G$). Inset is an optical image of the device. Scale bar is 10 µm. Metal lines are 1 µm wide. (c) Room temperature transfer characteristic ($I_D - V_G$) of the same device measured with $V_D = 400$ mV. The blue and red data curves correspond to $I_D$ displayed in log (left axis) and linear (right axis) scales, respectively.

Figure 2. (a) Two-probe electrical conductance ($G_{2p}$, log scale) and (b) Seebeck coefficient ($-S$) for device #1 (two-layer $MoS_2$) as a function of $V_G$ at six different temperatures ($T$) from 80 to 300 K. (c) Four-probe electrical conductance ($G_{4p}$ , log scale) as a function of $V_G$ in the ON state for device #1. (d) Arrhenius plot of $G_{4p}$ (log scale) vs. $1000/T$ for three different $V_G$'s showing a transition from metallic to insulating behaviors with decreasing $V_G$.

Figure 3. (a) The four-probe electrical conductivity ($\sigma_{4p}$) and (b) Seebeck coefficient ($-S$) of $MoS_2$ flakes of various thicknesses as functions of the back-gate voltage ($V_G - V_{th}$, relative to the threshold voltage) measured at room temperature. (c) $\sigma_{4p}$ and (d) $-S$ of $MoS_2$ as functions of the thickness (number of layers) measured at different $V_G - V_{th}$ values. These values correspond to a certain 2D charge carrier density inside the $MoS_2$ channel. Conductivity shows a maximum at two layers, while $-S$ shows a slight peak at four layers.

Figure 4. (a) Thermoelectric power factor ($PF$) versus the back-gate voltage ($V_G - V_{th}$, relative to the threshold voltage) for the different number of layers measured at room temperature. (b) $PF$ as a function of the number of layers measured at different $V_G - V_{th}$ values. $PF$ shows significant enhancement as the number of layers decreases down to two layers. The single-layer device has a low $PF$ mainly because of a low Seebeck coefficient (see Figure 3d).

Figure 5. Theoretically fitted (solid blue lines) (a) Seebeck coefficient ($-S$) and (b) four-probe electrical conductivity $\sigma_{4p}$ vs. $V_G - V_{th}$ for single-layer and double-layer $MoS_2$, plotted along with corresponding experimental data of Figure 3 (dashed red lines). $-S$ calculated from the Mott formula (dashed black line) for the single layer is also plotted in (a). (c) Seebeck coefficient ($-S$) vs. four-probe electrical conductivity ($\sigma_{4p}$) and (d) thermoelectric $PF$ vs. Fermi energy ($E_F - E_C$, with resect to the bottom of the conduction band, $E_C$) for single and double-layer $MoS_2$. Solid lines are theoretical results and dashed lines show experimental measurements.

Figure 6. (a) Full electronic band dispersions of single (blue) and double (dashed red) layers of $MoS_2$ calculated from DFT. The energy is measured relative to the bottom of the conduction band ($E_c$). (b) Density of states (DOS) and (c) number of modes as functions of $E - E_C$ for single and double-layer $MoS_2$.

Figure 7. (a) Seebeck coefficient ($-S$) and (b) four-probe sheet conductance ($G_{sheet}$) vs. $E_F - E_C$ for single-layer and double-layer $MoS_2$. Solid blue lines are theoretical fits.

Figure 8. (a) Calculated 2D carrier concentration ($n_{2D}$) vs. $E_F - E_C$ and (b) $E_F - E_C$ vs. $V_G - V_{th}$ for single and double-layer $MoS_2$. Note that $V_G - V_{th} = 0$ V corresponds to $E_F - E_C = -70$ meV and $-120$ meV for single layer and double layer, respectively. Calculated (c) mobility ($\mu$) and (d) average mean free path for single and double-layer $MoS_2$ as functions of $V_G - V_{th}$.

Figure 9. (a) Raman spectroscopy of single to four-layer $MoS_2$ flakes. Inset shows schematic atomic displacement of two Raman-active modes ($E_{2g}^1$ and $A_{1g}$) in the unit cell of the bulk $MoS_2$ crystal.[32,52] (b) AFM scanning of the six-layer $MoS_2$ flake showing a thickness of around 3.9 nm. The height profile was measured along the white horizontal line in the AFM image in the inset.

Figure 10. Schematics (not to scale) of (a) the current biasing four-probe and (b) voltage biasing two-probe electrical conductance/resistance measurements. The schematics were drawn for (a) with AC current and (b) with DC voltage, as used in our work.

Figure 11. (a) Arrhenius plot of $G_{4p}$ (log scale) vs $1/T$ for various $V_G$ values for the insulating part of Figure 2c. Solid lines are linear fits indicating thermally-activated transport for the high-temperature part. (b) Activation energy ($E_a$) as a function of $V_G$.

Figure 12. The four-probe electrical conductance in 23-layer $MoS_2$ with Al contacts showing a metal ($V_G > \sim -15$ V) to insulator ($V_G < \sim -15$ V) transition.

Figure 13. The measurement set-up (schematic not to scale) used for Seebeck coefficient in AC (with AC $I_{Heater}$) and DC (with DC $I_{Heater}$) modes. If the measured open-circuit voltage ($V_{Open-Circuit}$) is caused by thermoelectric effects then $V_{Open-Circuit} = V_{thermal}$.

Figure 14. (a) The temperature rise at each thermometer as a function of $I_{Heater}^2$ for device #1 at room temperature. (b) Room temperature $V_{thermal}$ as a function of $I_{Heater}^2$ for two different back-gate voltages in the ON state. (c) Amplitude of the temperature in the device geometry presented in the inset of Figure 1b calculated from a finite-element simulation (using COMSOL) for $I_{Heater} = 4.8\,mA$. The simulated temperature difference across the $MoS_2$ channel is 12% lower than that we measure experimentally.

Figure 15. (a) The four-probe 2D sheet conductance ($G_{sheet}$) of $MoS_2$ flakes of various thicknesses as a function of the back-gate voltage ($V_G - V_{th}$, relative to the threshold voltage) measured at room temperature. (b) $G_{sheet}$ of $MoS_2$ as a function of the thickness (number of layers) measured at different $V_G - V_{th}$ values. (c) $\sigma_{4p}$ as a function of the thickness for $V_G = 0$ V. (d) The highest filed effect mobility (right) and $G_{sheet}$ at $V_G = 0$ V (left) of our devices vs. thickness. (e,f) $-S$ (e) and $PF$ (f) as functions of thickness for $V_G = 0$ V.

Figure 16. (a) The four-probe electrical conductivity ($\sigma_{4p}$) and Seebeck coefficient ($-S$) of another double-layer $MoS_2$ as functions of the back-gate voltage ($V_G - V_{th}$, relative to the threshold voltage) measured at room temperature. (b) $PF$ of the same sample vs $V_G - V_{th}$ showing $PF$ values as large as $\sim 53\,\frac{\mu W}{cmK^2}$. The large value of $PF$ achieved here compared to the double-layer $MoS_2$ device shown in the main text Figure. 4 is because in this device we were able to apply larger $V_G$'s.

Figure 17. (a) The two-probe resistance ($R_{2p}$, left) and conductance ($G_{2p}$, right) of the $MoS_2$ flake as functions of $V_G$. Inset shows the optical image of the thermoelectric device. (b) $V_{open-circuit}$ across the channel measured as a function of $I_{Heater}$ deep into the OFF state of the device. No parabolic behavior is observed in this region. (c) $V_{open-circuit}$ across the channel measured as a function of $I_{Heater}$ close to the onset of ON state (blue curve on the left axis) and deep into the ON state of the same device (red curve on the right axis). (d) Gate-dependent $V_{open-circuit}$ in the OFF state for $I_{Heater} = 4, 0, -4$ mA.

Figure 18. Schematics (not to scale) of the device showing the resistive coupling (large but finite resistance) from the gold electrodes and heater to the conducting silicon back gate through the $SiO_2$.

Figure 19. $V_{open-circuit}$ across the channel as functions of (a) $I_{Heater}$ and (b) $V_G$ in the OFF state measured with the SPA. The voltage measured when the heater is OFF is due to the leakage current from either the back-gate voltage or the nanovoltmeter.

Figure 20. (a) A comparison between AC measurements when the lock-in amplifier is directly used to measure $V_{open-circuit}$ (dashed lines) and when a SR560 pre-amplifier is used to reduce the loading effects. (b) The phase shift between the heater current and $V_{open-circuit}$ as a function of the back-gate voltage for three different frequencies when the lock-in amplifier is directly used. The phase shift deviates from 90 deg as the device goes into the subthreshold regime. Shaded regions in both plots indicate that thermoelectric measurements would be unreliable particularly for lock-in frequencies higher than ~ 6 Hz.

Figure 1.

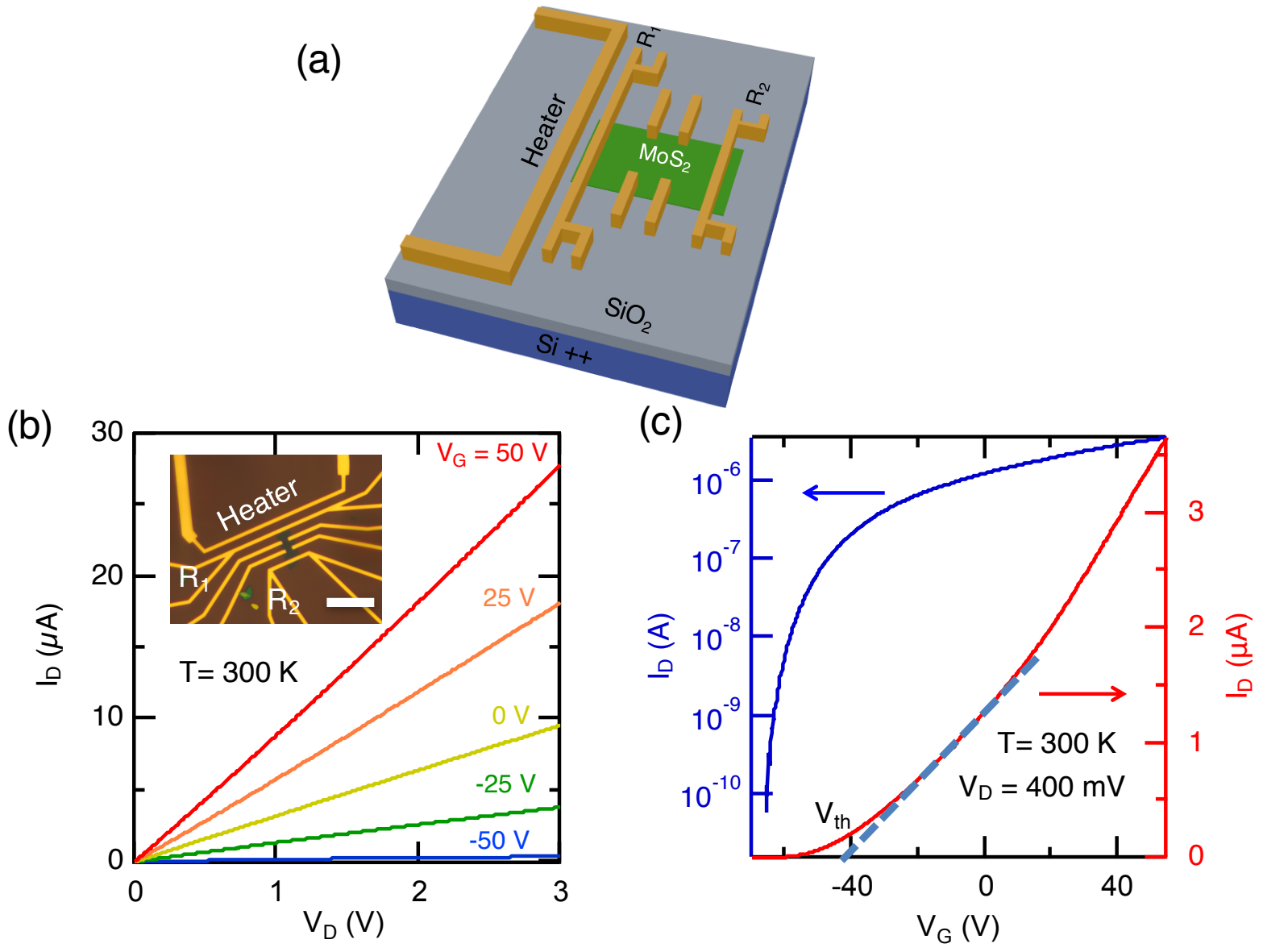

Figure 2.

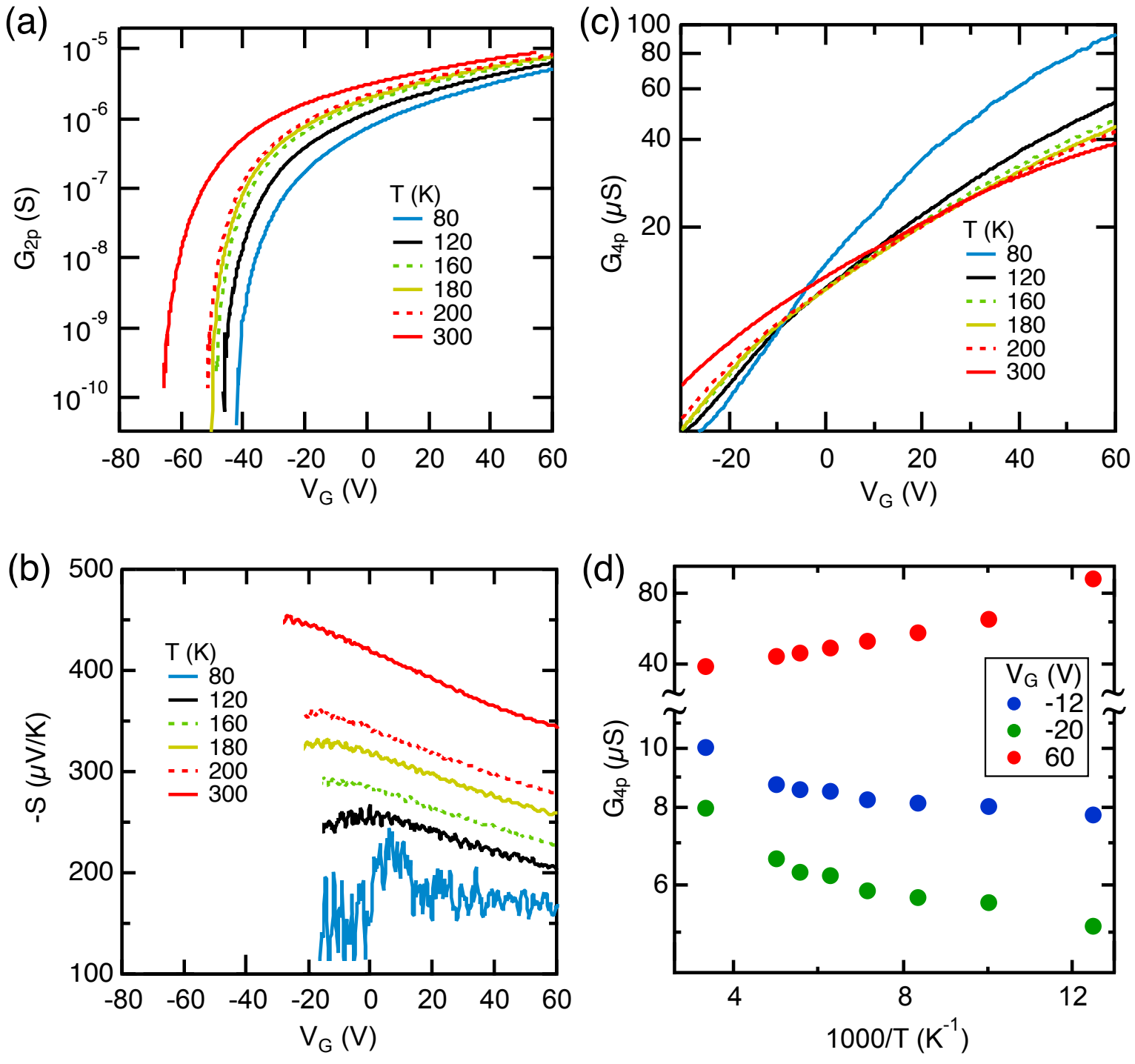

Figure 3.

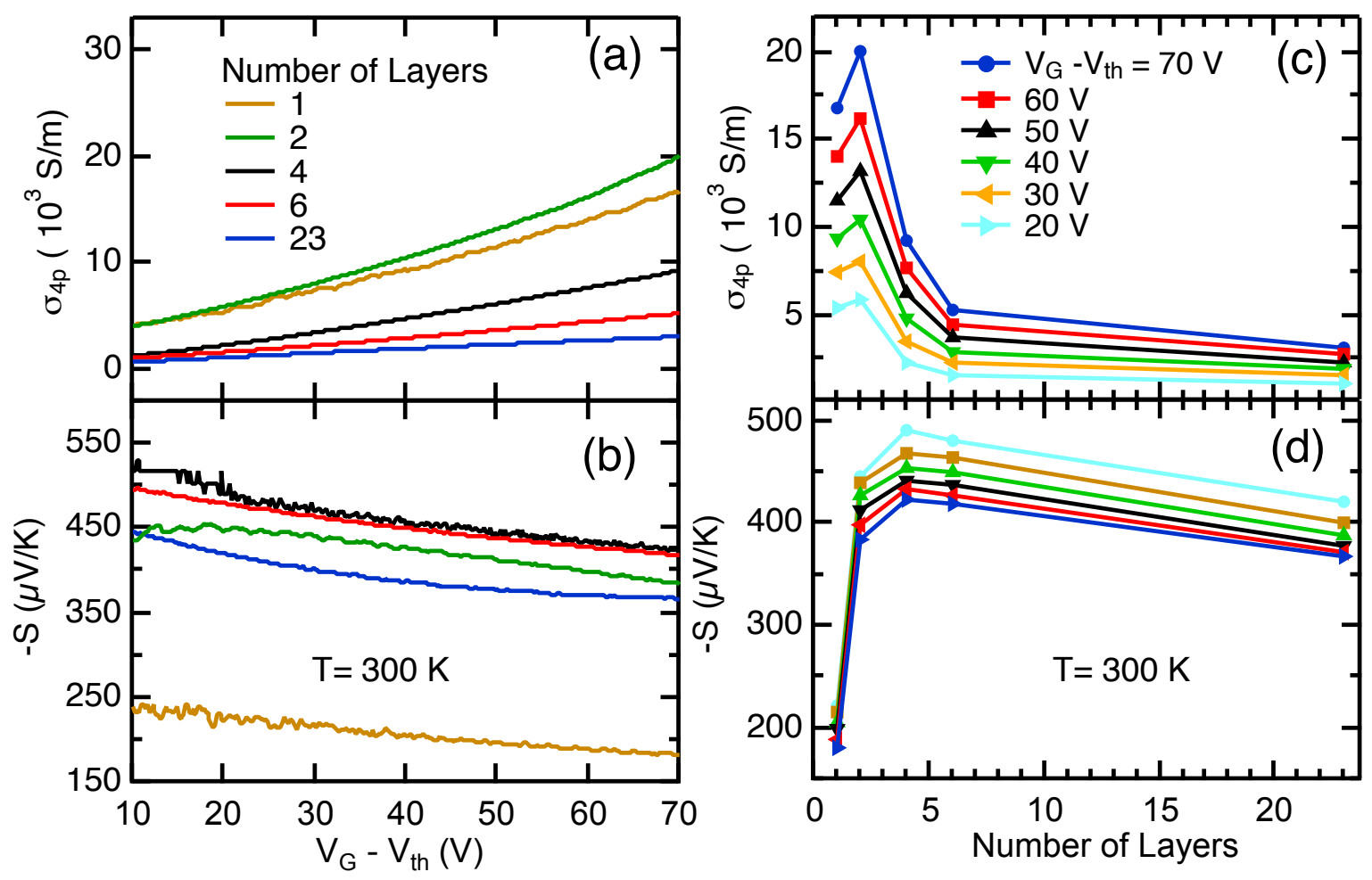

Figure 4.

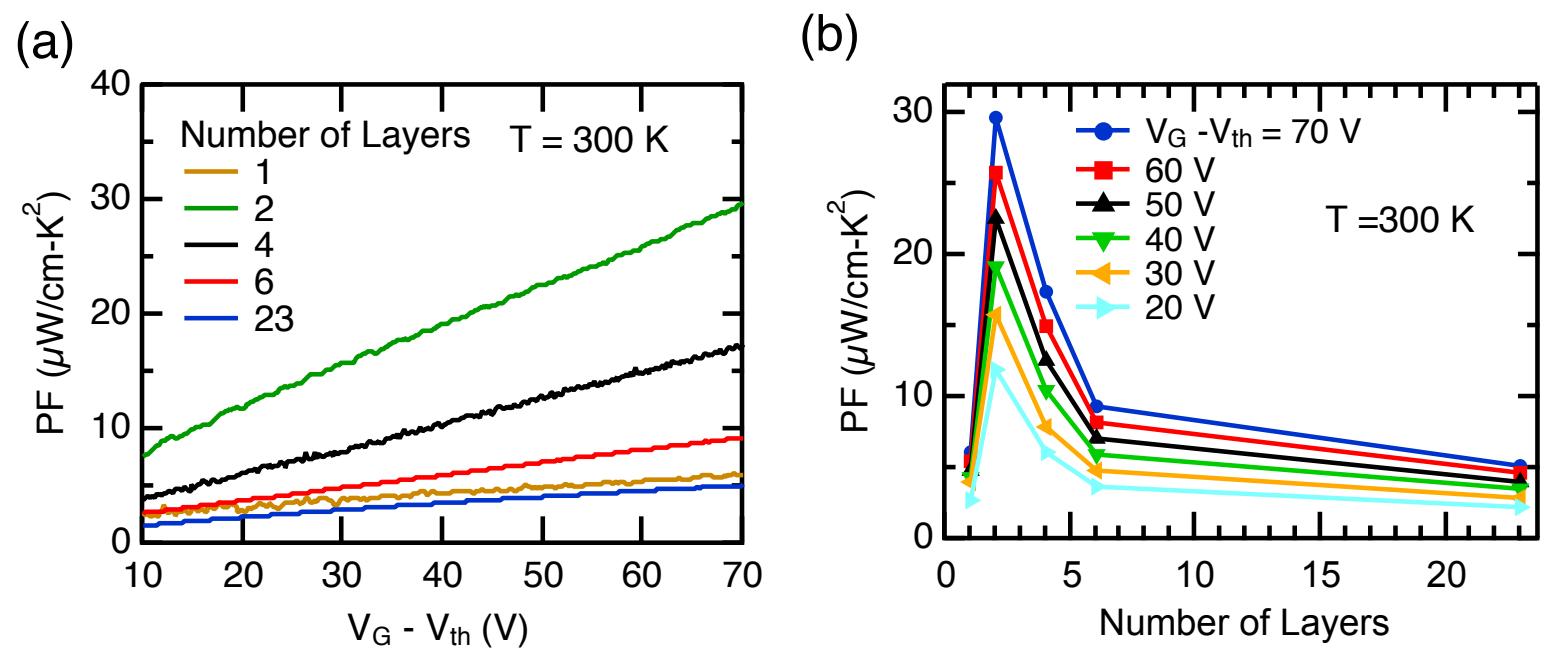

Figure 5.

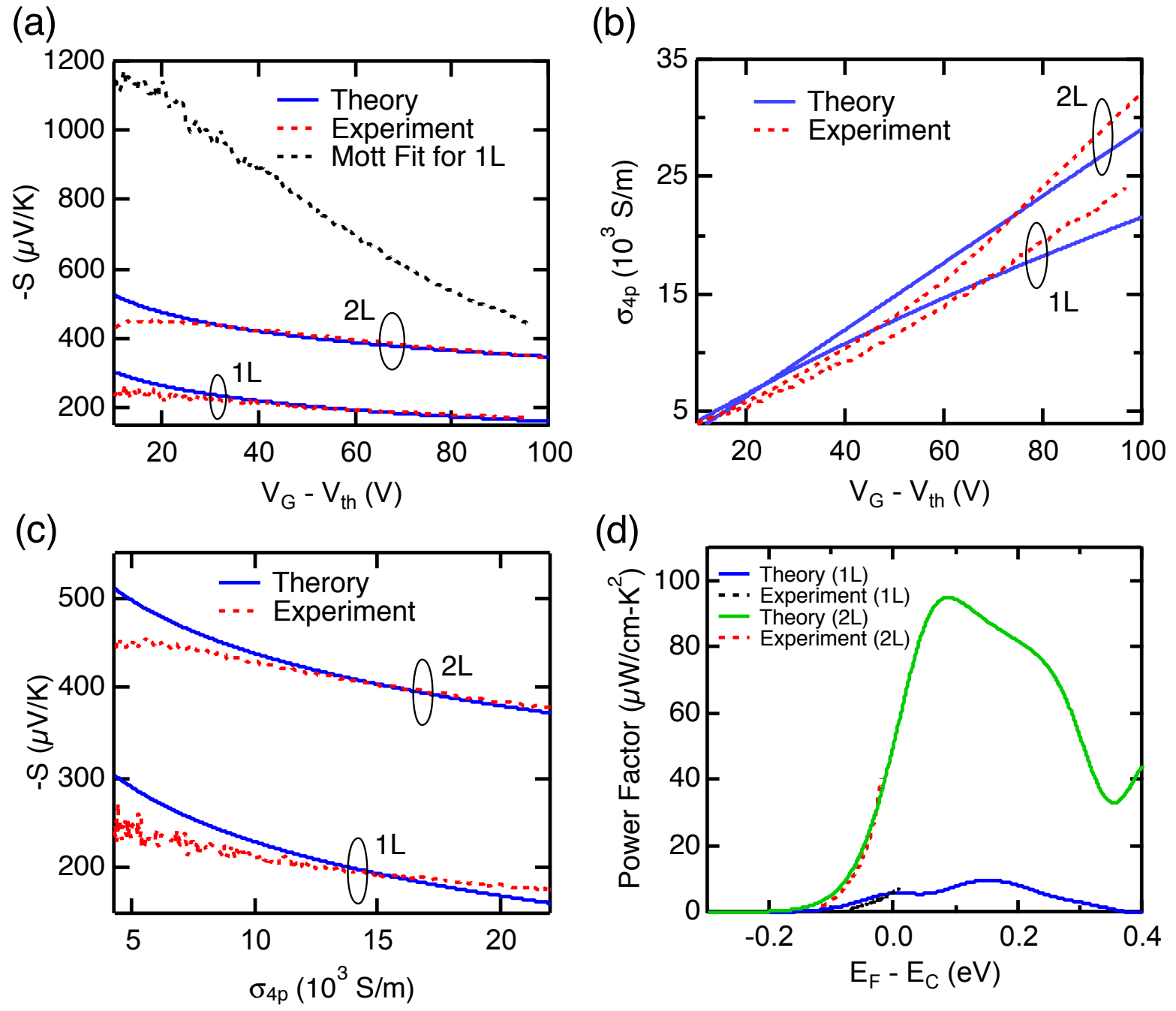

Figure 6.

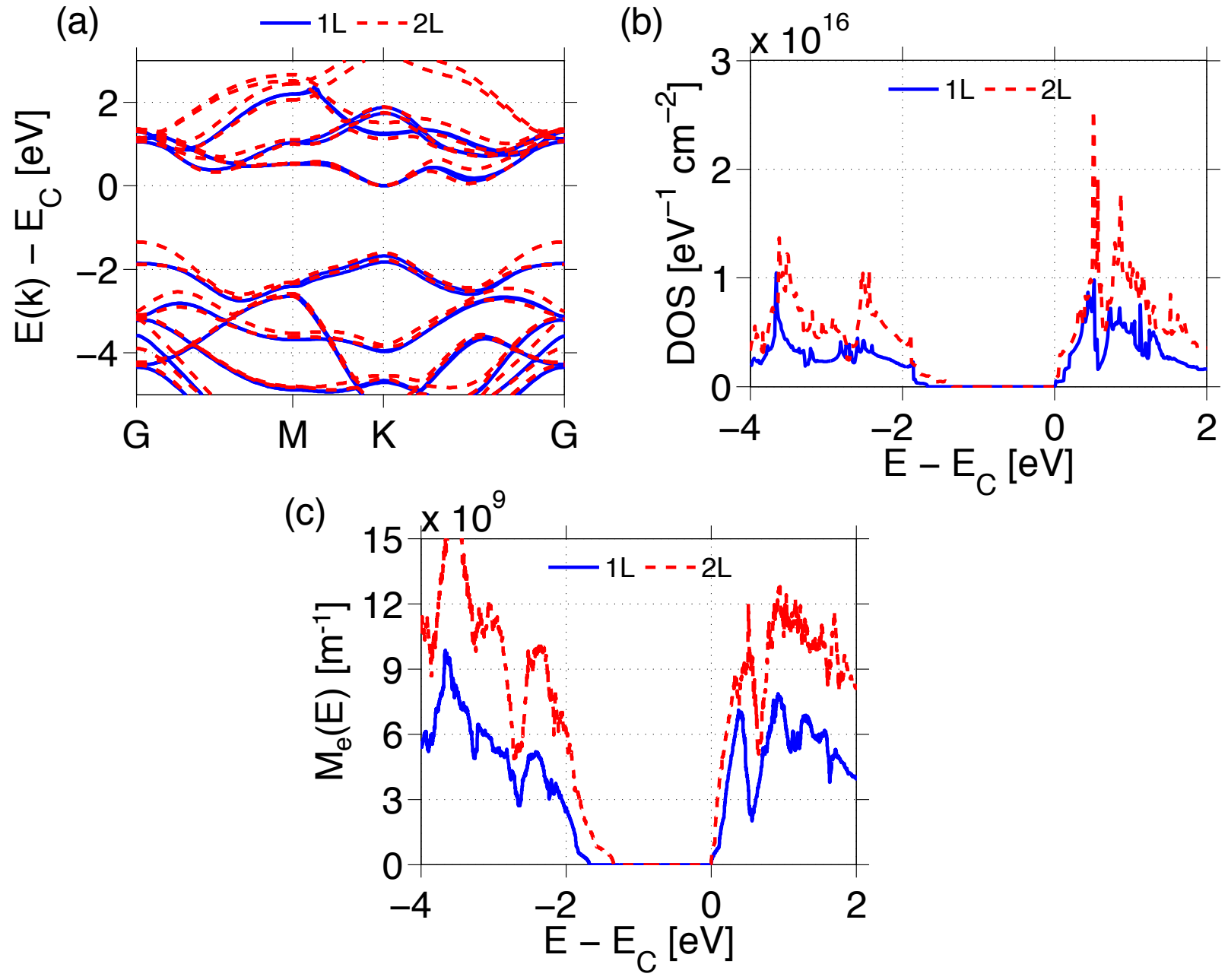

Figure 7.

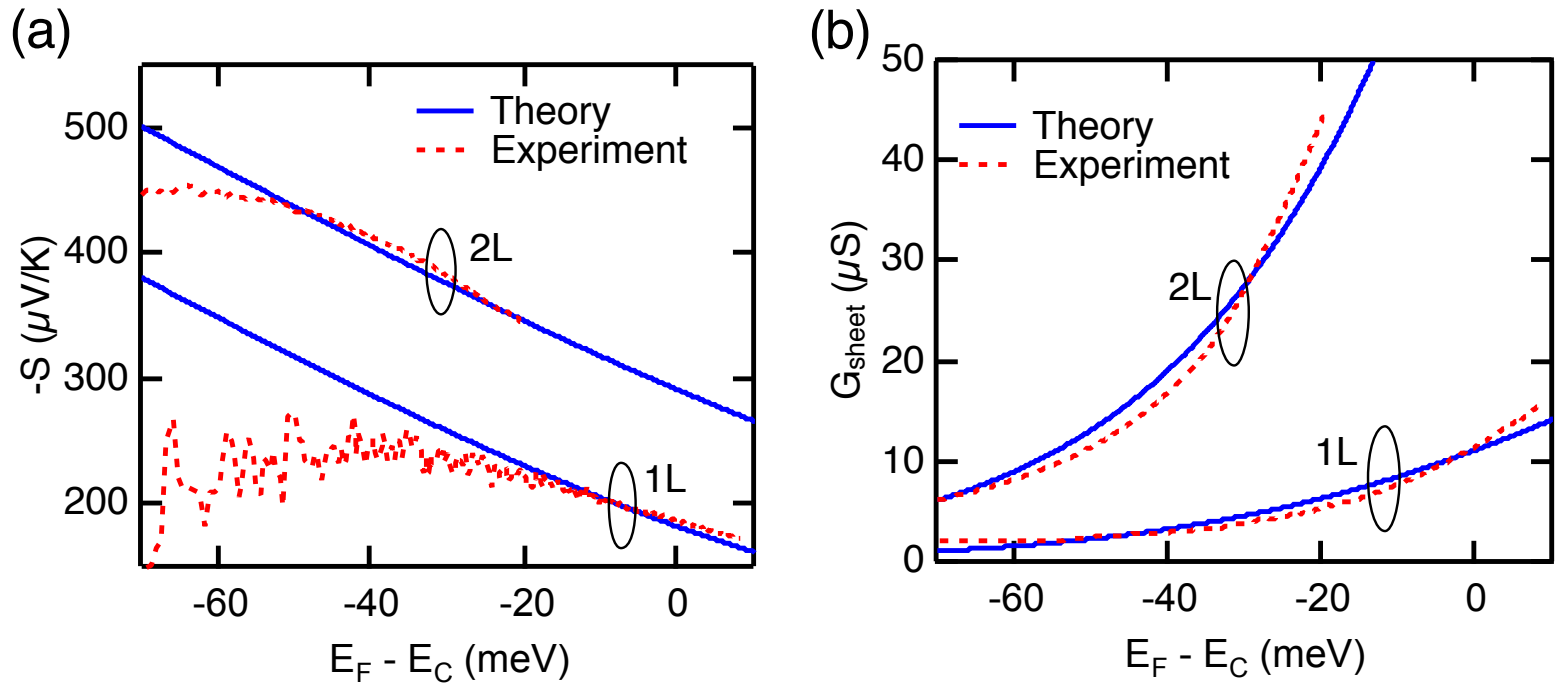

Figure 8.

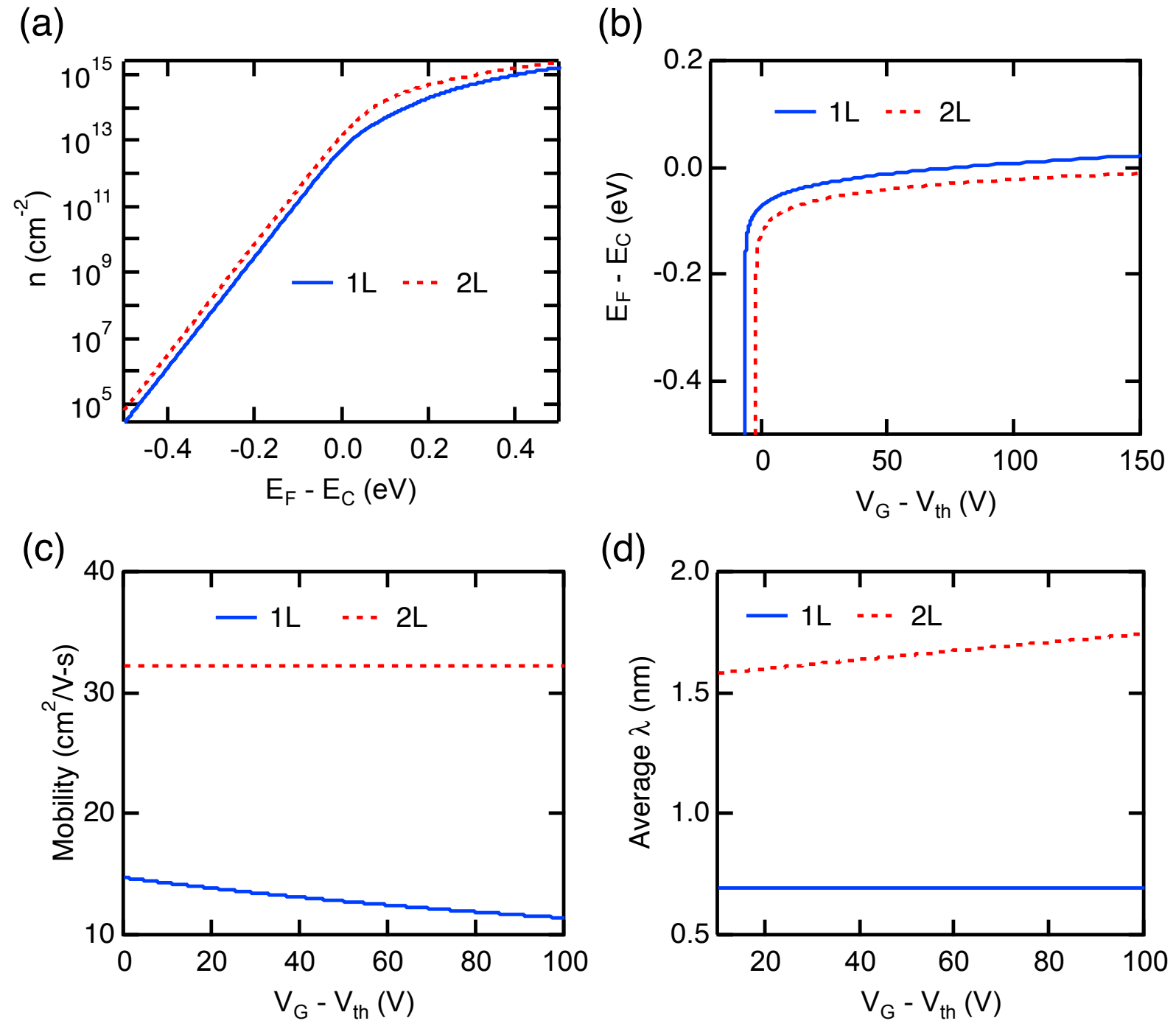

Figure 9.

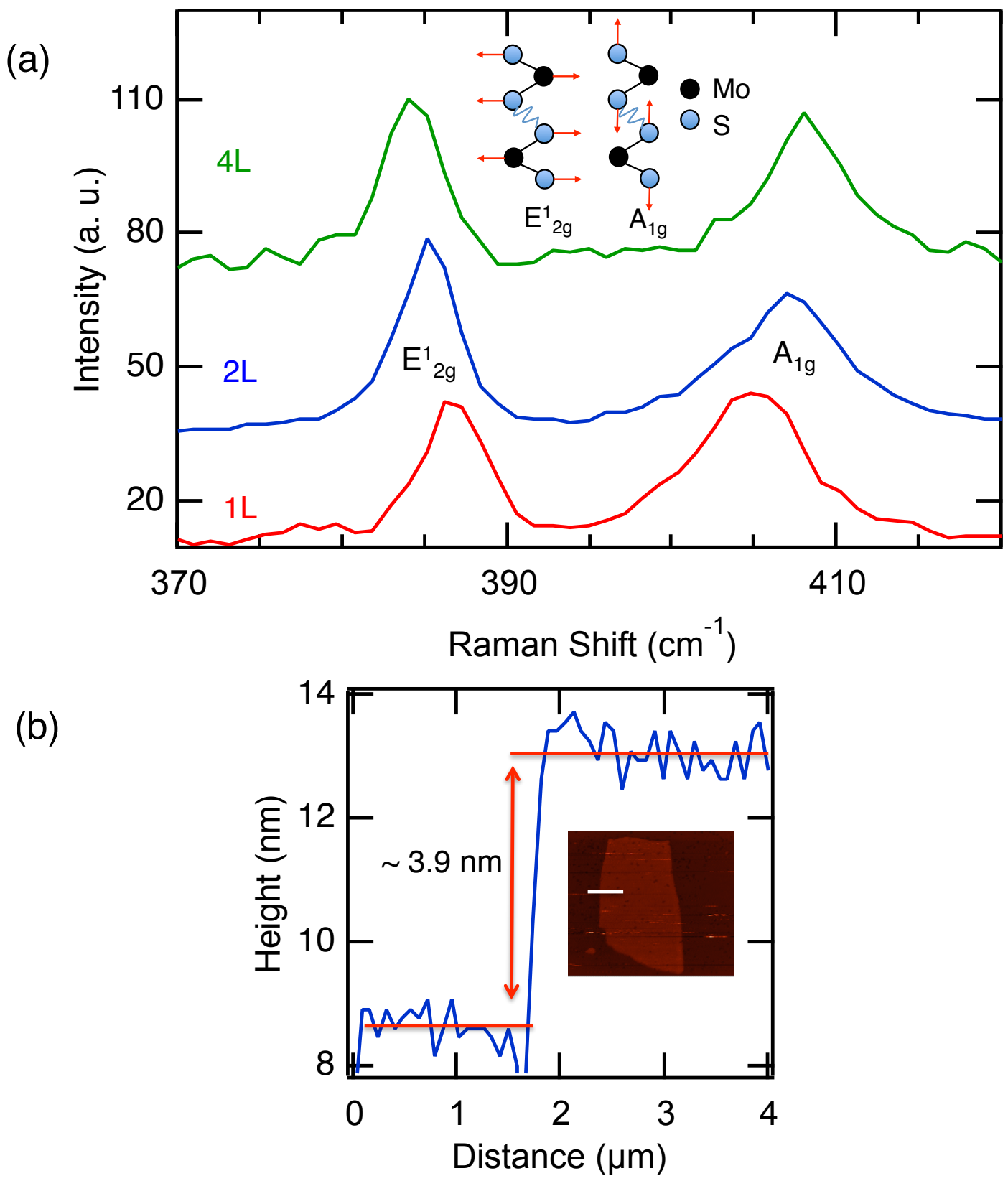

Figure 10.

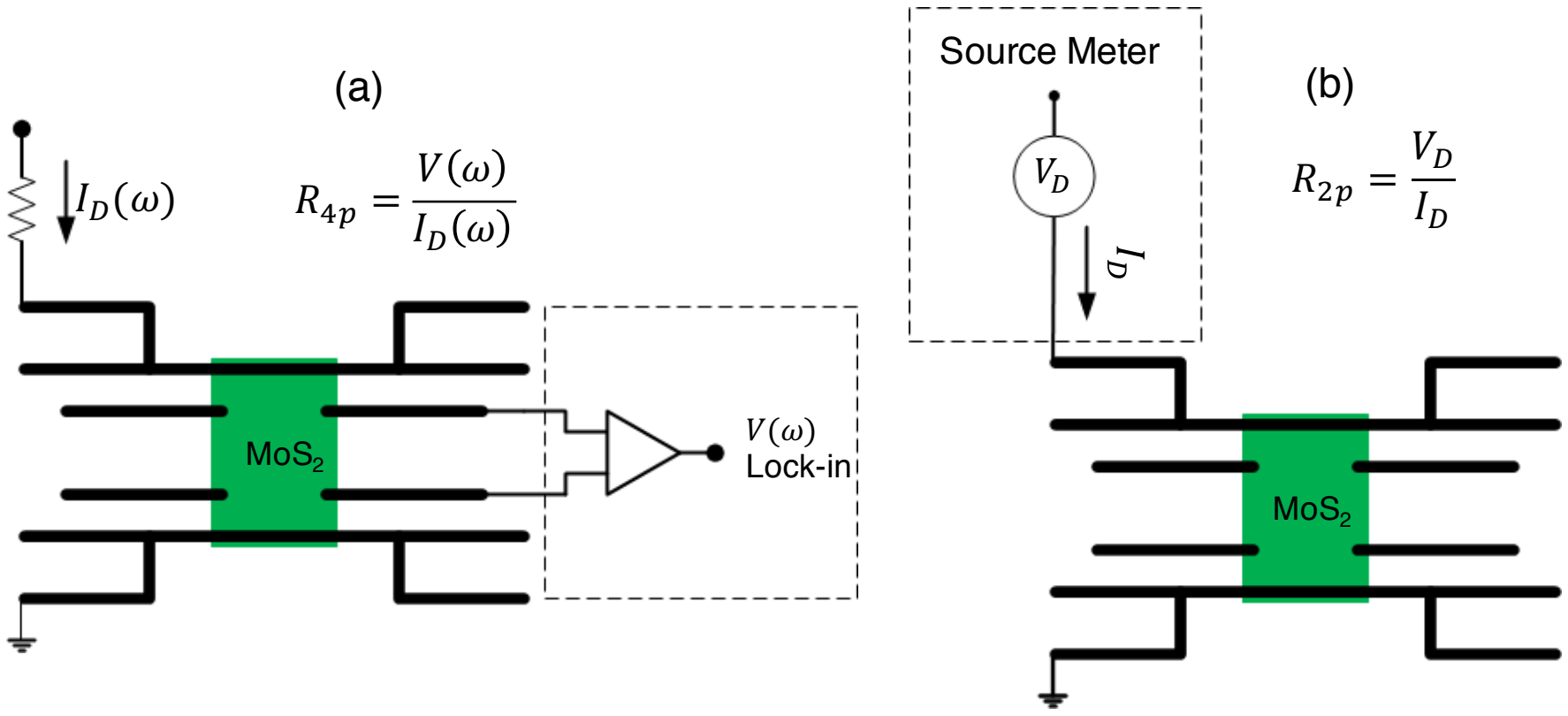

Figure 11.

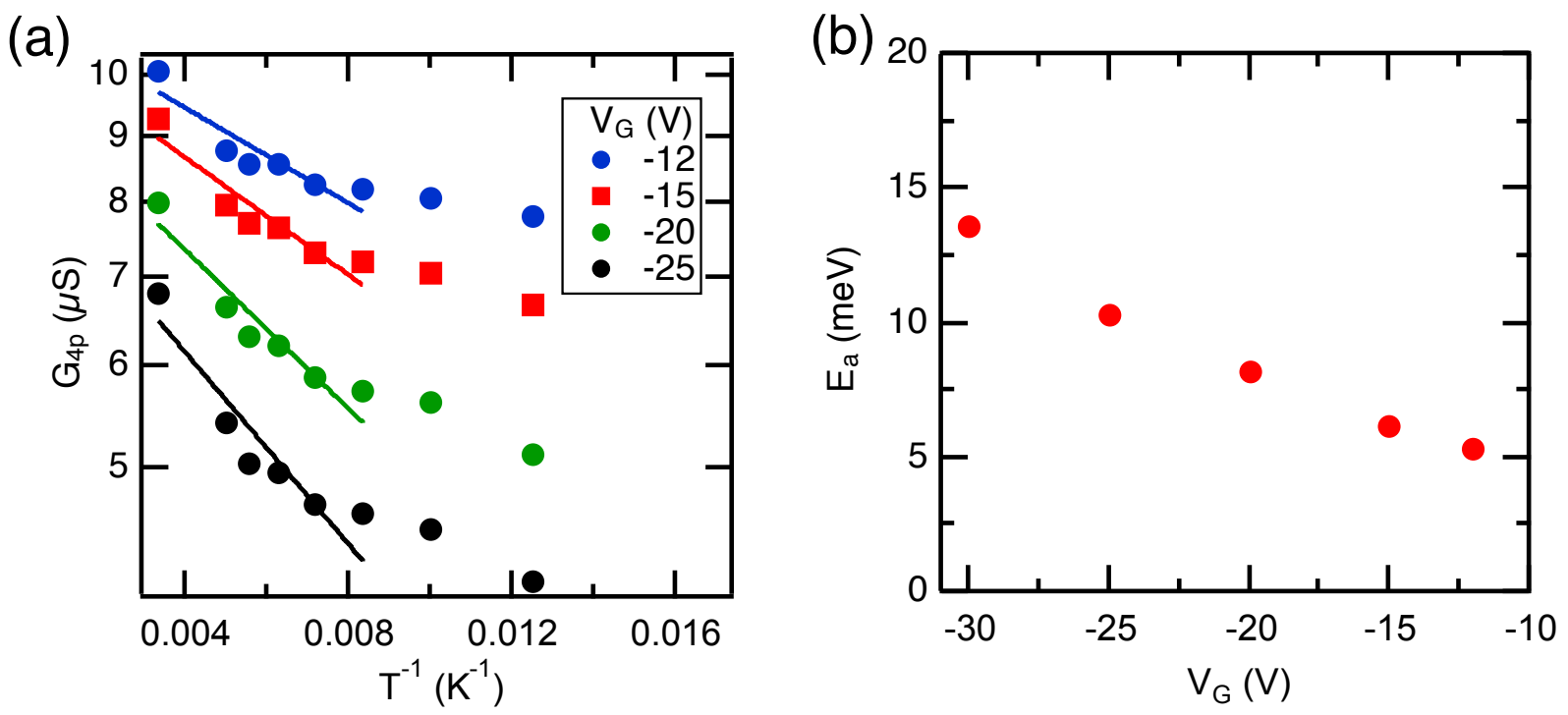

Figure 12.

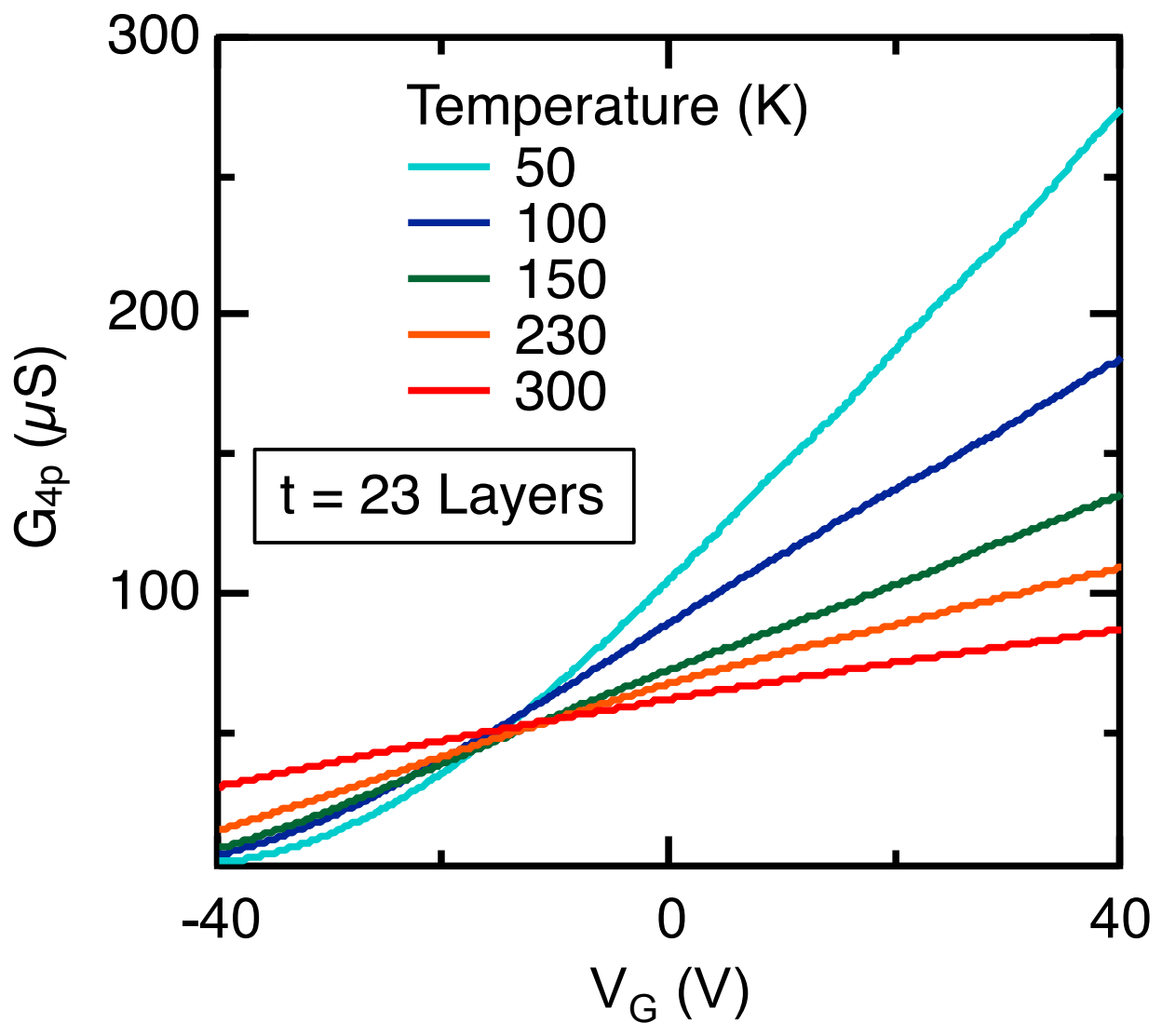

Figure 13.

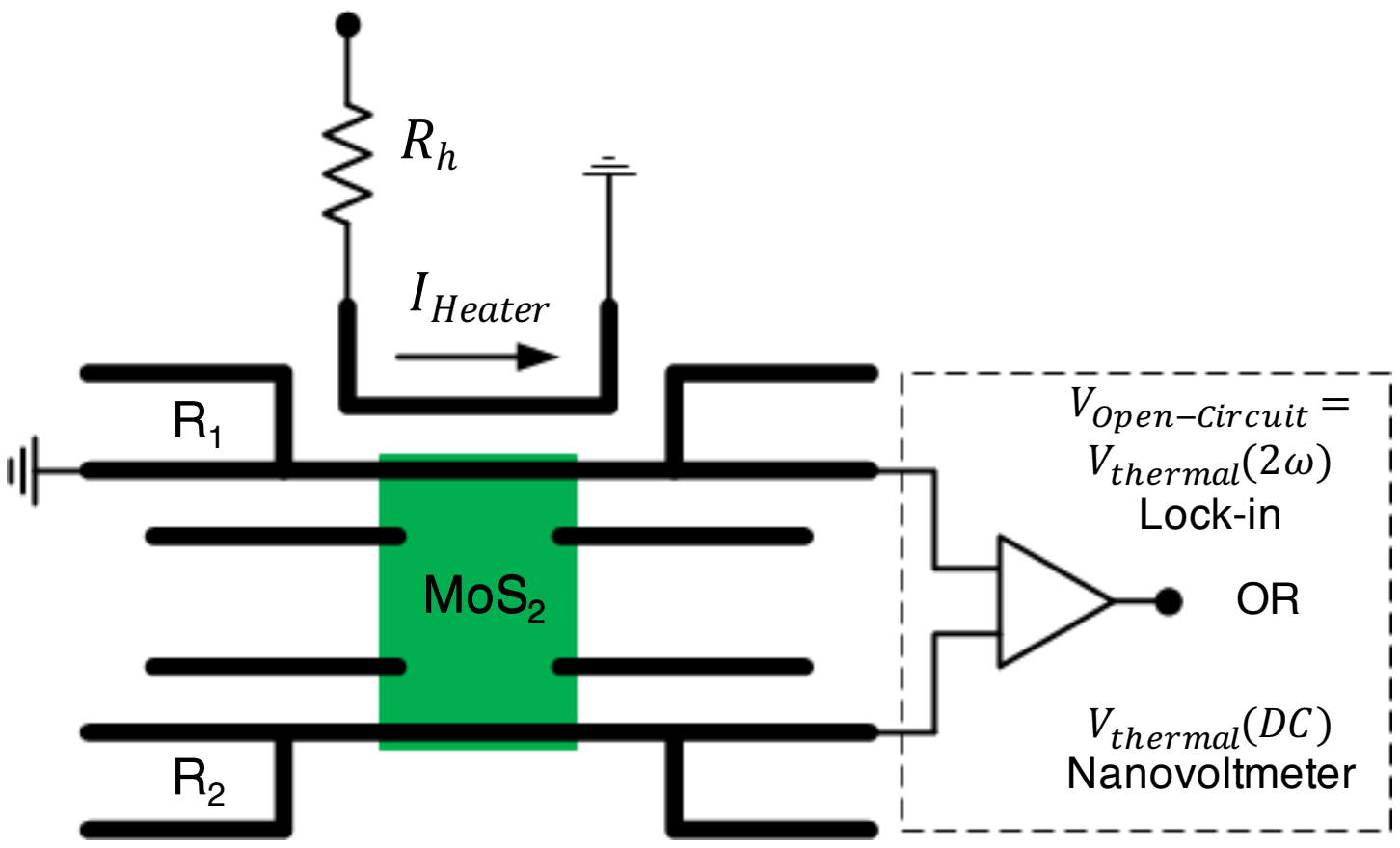

Figure 14.

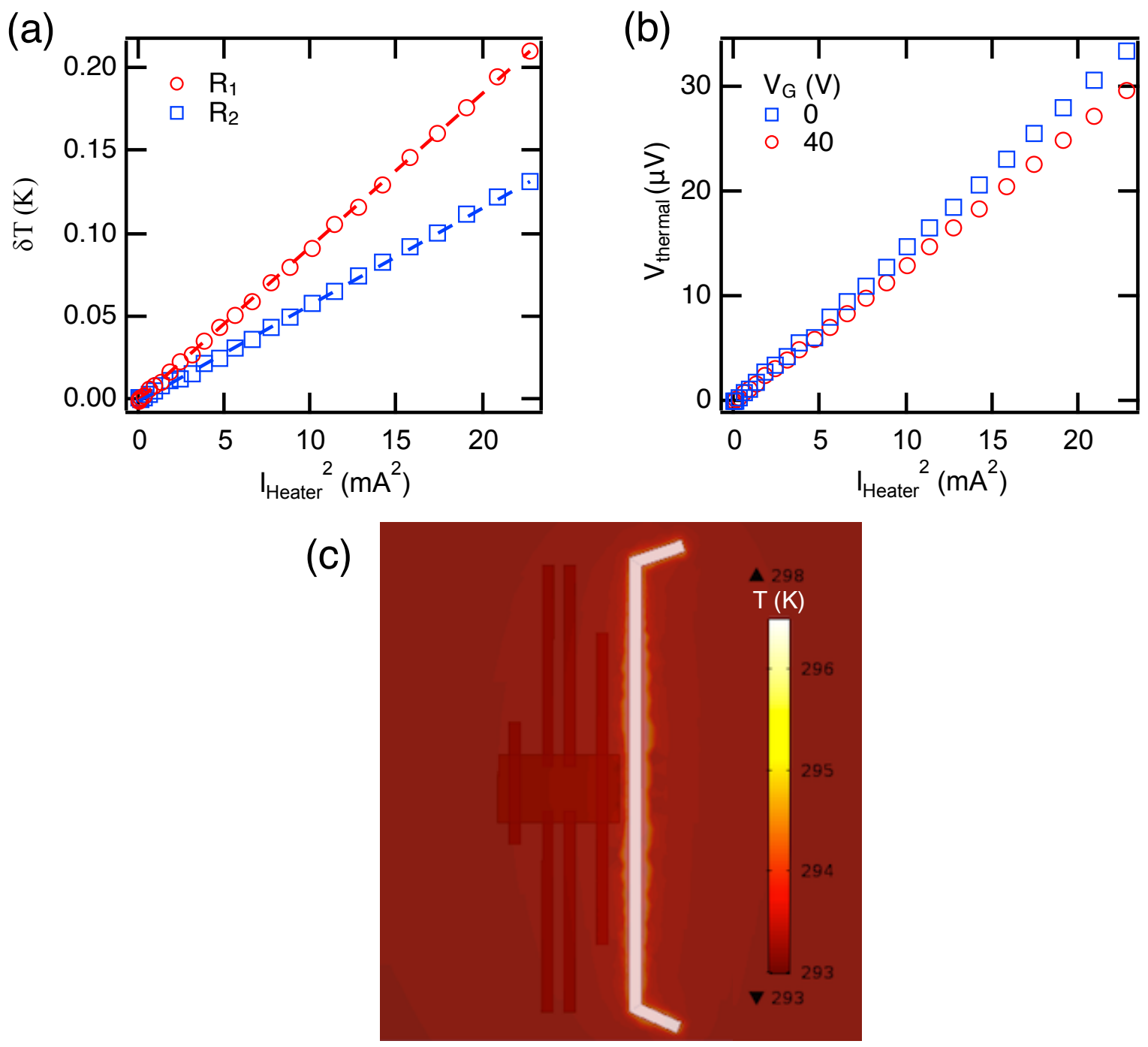

Figure 15.

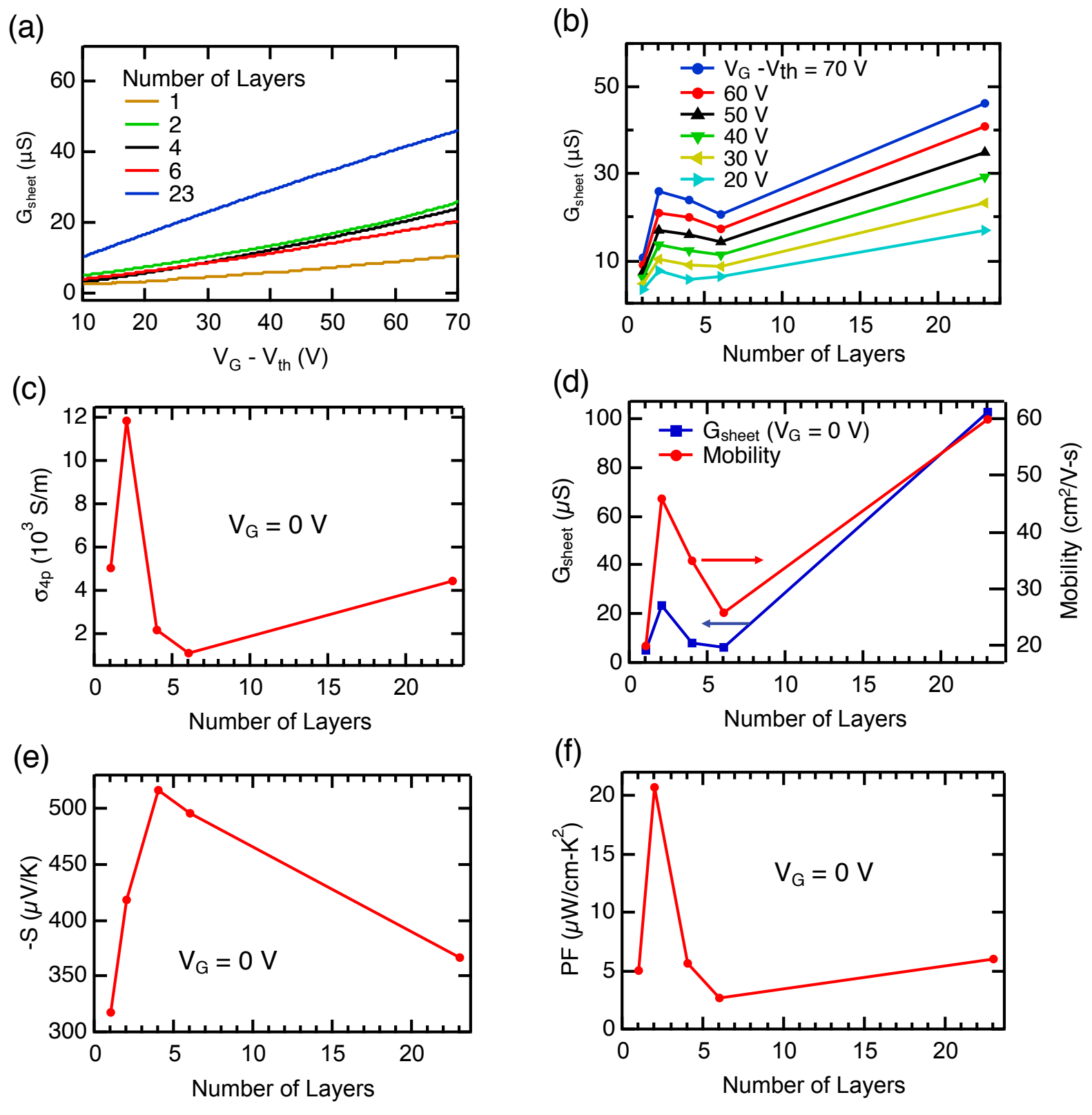

Figure 16.

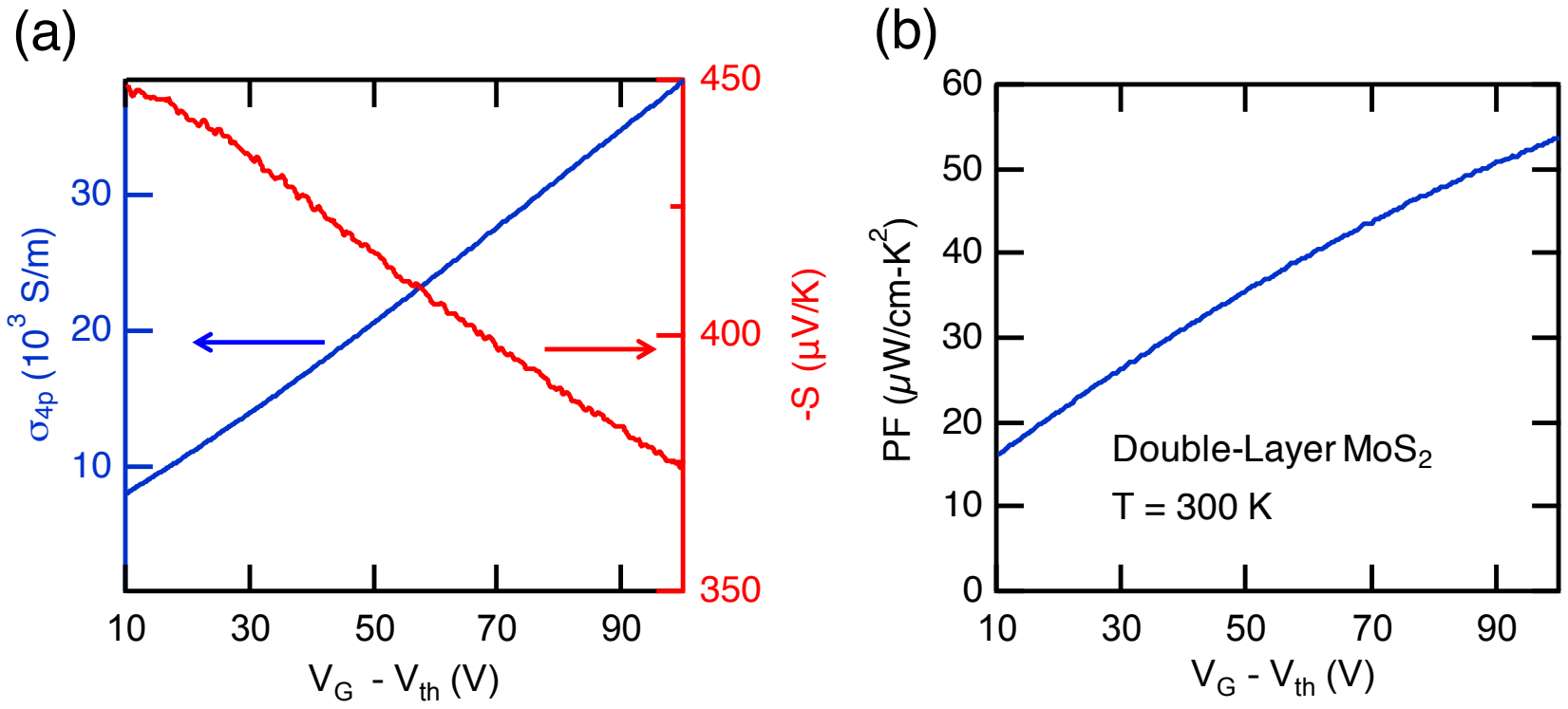

Figure 17.

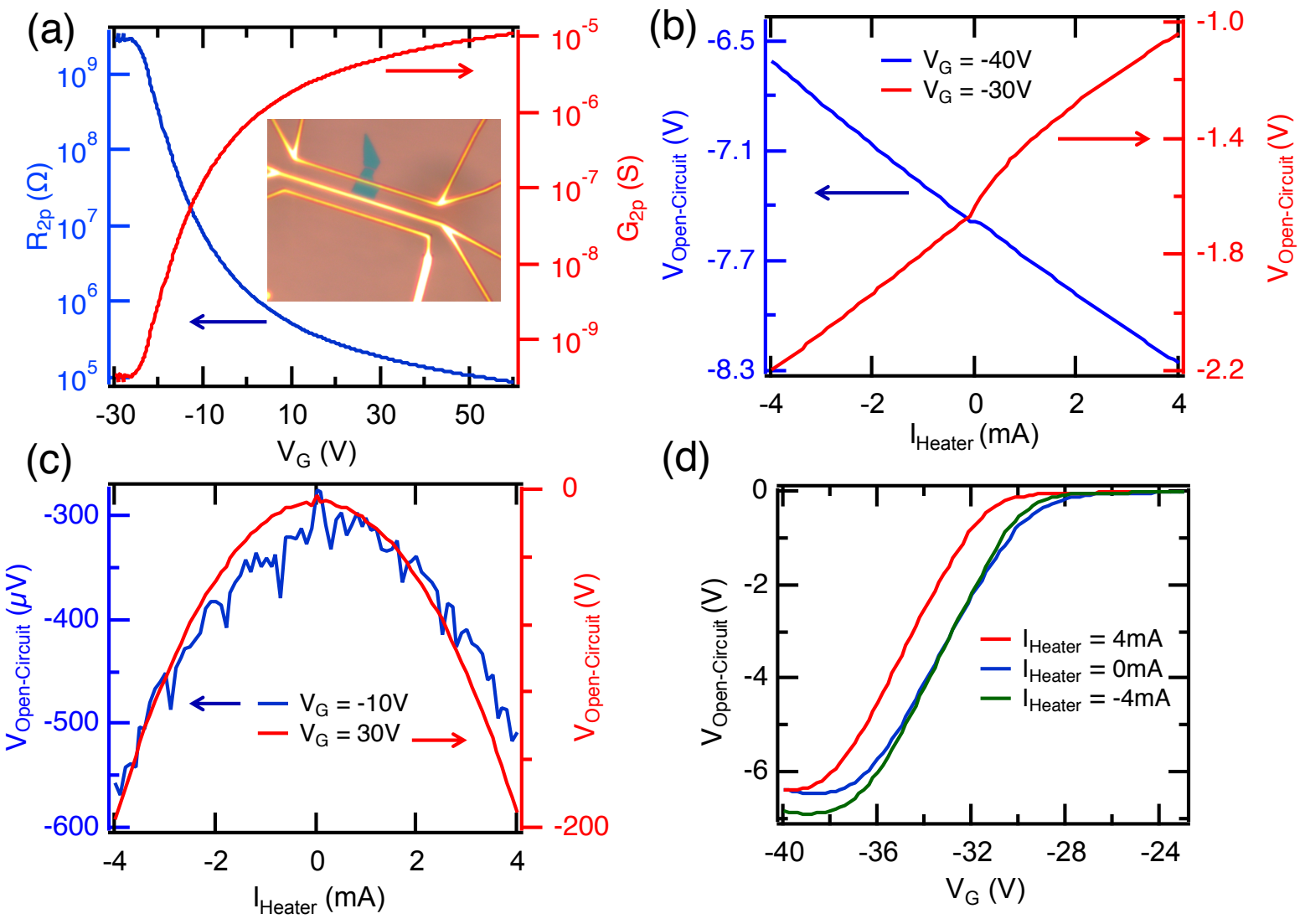

Figure 18.

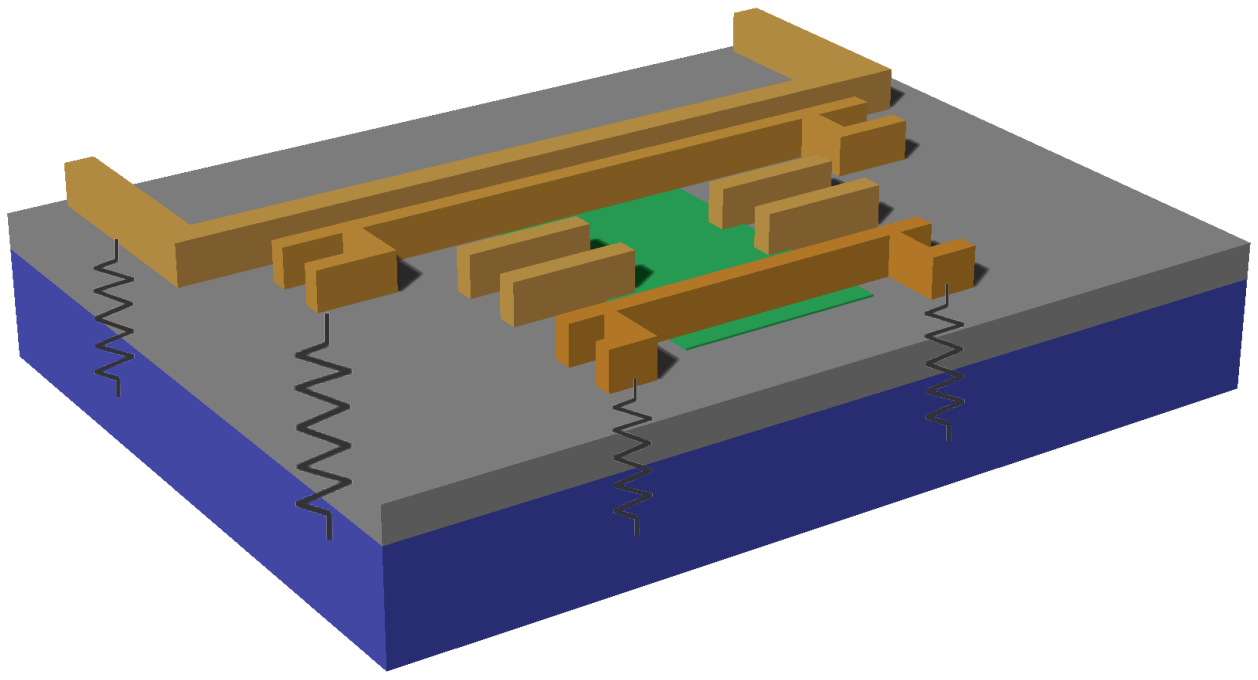

Figure 19.

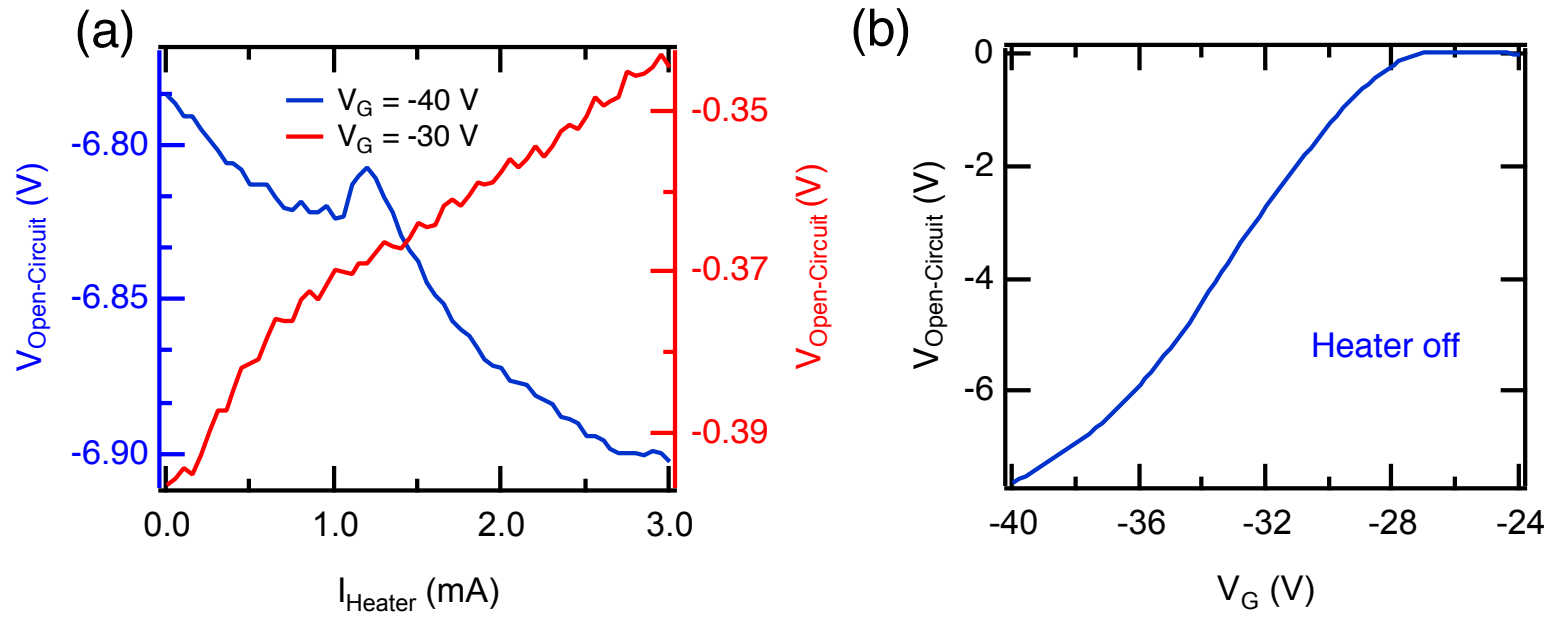

Figure 20.

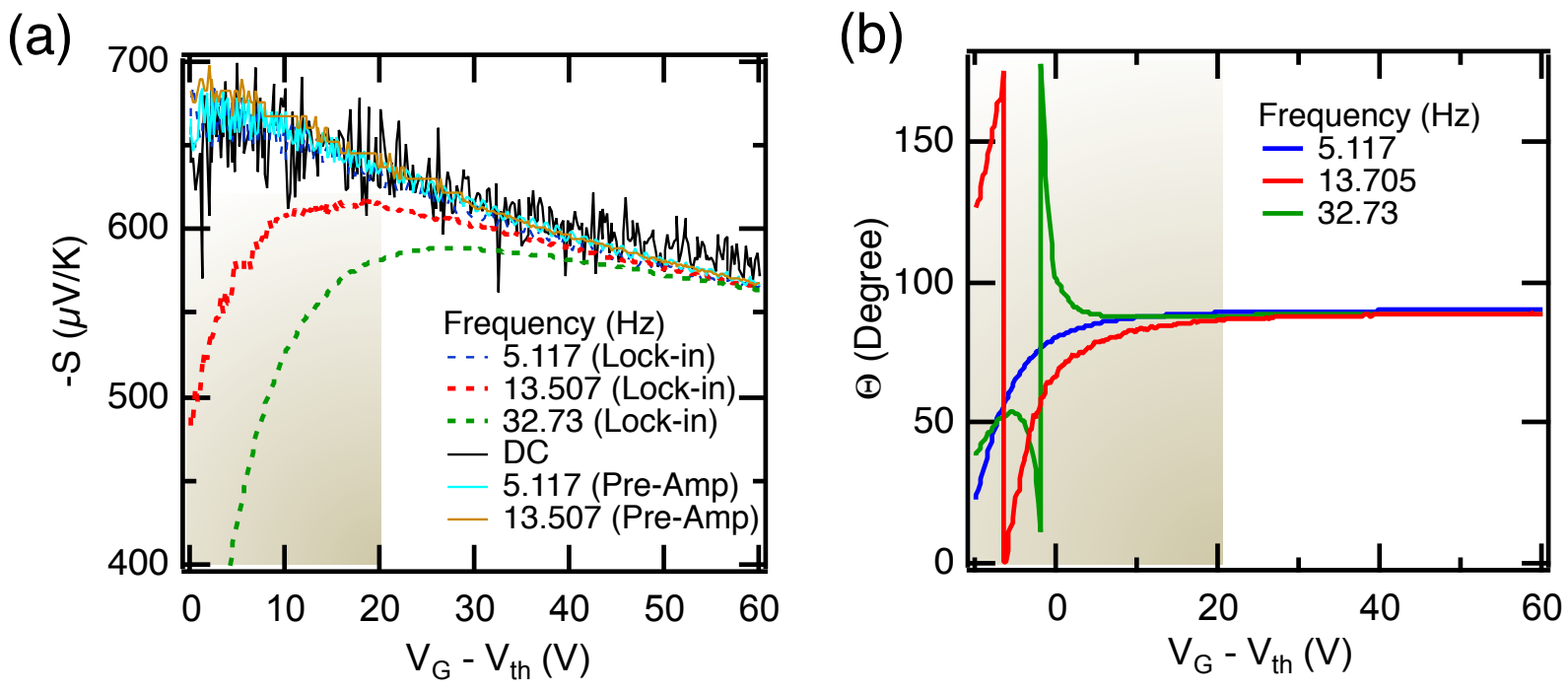